\documentclass[12pt,preprint]{aastex}

\usepackage{ulem}

\usepackage{epsfig}

\begin{document}

\title{Evolution and Distribution of Magnetic Fields from AGNs in Galaxy Clusters. I. The Effect
  of Injection Energy and Redshift }


\author{Hao Xu\altaffilmark{1}, 
Hui Li\altaffilmark{1},  
David C. Collins\altaffilmark{2},  
Shengtai Li\altaffilmark{1},
and Michael L. Norman\altaffilmark{2}}
\altaffiltext{1}{Theoretical Division, Los Alamos National Laboratory, Los
  Alamos, NM 87545; hao\_xu@lanl.gov, hli@lanl.gov, sli@lanl.gov}
\altaffiltext{2}{Center for Astrophysics and Space Sciences,
  University of California, 
San Diego, 9500 Gilman Drive, La Jolla, CA 92093; dcollins@physics.ucsd.edu, mlnorman@ucsd.edu}

\begin{abstract}
   We present a series of cosmological magnetohydrodynamic (MHD) simulations that simultaneously follow
   the formation of a galaxy cluster and evolution of magnetic fields ejected by an Active 
    Galactic Nucleus (AGN). Specifically, we investigate the influence of both the epoch of 
   AGN (z $\sim$ 3-0.5) and the AGN energy ($\sim$ 3 $\times$ 10$^{57}$ - 2 $\times$ 10$^{60}$ ergs)
   on the final magnetic field distribution in a relatively massive cluster (M$_{vir}$ $\sim$10$^{15}$ M$_\odot$).
  We find that as long as the AGN magnetic fields are ejected before the major mergers in the 
  cluster formation history,  magnetic fields can be transported throughout the
  cluster and can be further amplified by the
  intra-cluster medium (ICM) turbulence cause by hierarchical mergers during the cluster
  formation process. The total magnetic energy in the cluster
  can reach $\sim$ $10^{61}$ ergs, with micro Gauss fields distributed over $\sim$ Mpc scale. 
  The amplification of the total magnetic energy by the ICM turbulence can be significant, 
 up to $\sim$1000 times in some cases. 
  Therefore even weak magnetic fields from AGNs can be used to magnetize the cluster to the observed level. 
  The final magnetic energy in the ICM is determined by the ICM 
  turbulent energy, 
  with a weak dependence on the AGN injection energy. We discuss the properties of
  magnetic fields throughout the cluster and the synthetic Faraday rotation 
  measure maps they produce. We also show that high spatial resolution over most of the
  magnetic regions of the cluster is very important to capture the small scale dynamo process and
  maintain the magnetic field structure in our simulations.

\end{abstract}

\section{Introduction}

There is growing evidence that the intracluster medium (ICM) of galaxy clusters is permeated with magnetic
fields, as indicated by the detection of large-scale, diffused radio
emission called radio halos and relics \citep[e.g.][]{Carilli02, Ferrari08, Giovanini09}.
The radio halos 
are sometimes extended over $\ge 1$ Mpc, covering the whole cluster. By assuming that
the magnetic energy is comparable to the
total energy in relativistic electrons, one often deduces that 
the magnetic fields in the cluster halos  
can reach $0.1-1.0$ $\mu$G and the total magnetic energy can be
as high as $10^{61}$ ergs \citep{Feretti99}. 

The Faraday rotation measurements (FRM) are extensively used to study the magnetic fields in 
galaxy clusters. 
By combining FRM and density measurements, magnetic field strengths in
clusters have been measured as high as a few to ten $\mu$G level, mostly in
the cluster core region\citep{Carilli02}. More interestingly, FRM was used to suggest that magnetic
fields can have a Kolmogorov-like turbulent spectrum in the cores of
clusters \citep{Vogt03}, with an energy spectrum peak at several kpc. Other studies \citep{Eilek02, Taylor93, Colgate00} 
have suggested that the coherence scales of magnetic fields can range from a 
few kpc to a few hundred kpc, implying
large amounts of magnetic energy and fluxes in the ICM. Recently,
radial profiles of magnetic fields extended to the outer part of clusters
 have been estimated with FRM from observations of additional radio galaxies behind these clusters
 \citep{Govoni06,Guidetti08,Bonafede10}. 
 To accurately constrain the magnetic
fields in clusters, knowledge of the three dimensional distribution of the magnetic
fields is required.  This will be especially true when  the Extended
Very Large Array (EVLA) becomes operational, as it will be able to obtain FRM
over large areas of galaxy clusters. It is also suggested that the evolution 
and distribution of magnetic fields in clusters may be important to the cluster formation because they
play a crucial role in
processes such as heat transport, which consequently affect the
applicability of clusters as sensitive probes for cosmological
parameters \citep{Voit05}. In addition, the distribution of magnetic fields 
can potentially serve as a tracer of the merger history of a cluster.

Magnetic field evolution is complicated and difficult to study analytically in 
highly non-linear systems, such as galaxy clusters, and therefore
cosmological MHD simulations are  used to study the properties of magnetic fields in the ICM.
Although the existence of cluster-wide magnetic fields is clear, their
origin is still poorly understood. Consequently, initial magnetic fields are
usually added to the simulations by hand. Simulations have been performed with initial magnetic fields 
either from some random or uniform fields at high redshifts \citep{Dolag02, Dubois08, Dubois09} 
or from the outflows of normal galaxies \citep{Donnert08}. 
These simulations found that fields can be further amplified by cluster mergers
\citep{Roettiger99} and turbulence. Though it is not clear whether
the initial conditions of their simulations are correct, their findings roughly match results from observations.
This suggests that cosmological MHD simulations are a good way to study cluster magnetic fields.
On the other hand, very small seed fields from some first principle mechanism,
like Biermann battery effect have also been studied \citep{Kulsrud97,Xu09b}. It
has been suggested that very week seed fields can be amplified by dynamo
processes in clusters \citep{Ryu08}, though current simulations have not been able to model such processes 
self-consistently. In addition, the exact
mechanism for dynamo is still being debated \citep[e.g.][]{Bernet08}.

Observationally, large scale radio jets and lobes from AGNs serve as one very intriguing
source of cluster magnetic fields, because they could carry large amounts
of magnetic energy and flux
\citep{Burbidge59,Kronberg01,Croston05,McNamara07}.  The magnetization
of the ICM and the wider inter-galactic medium (IGM) by AGNs has been
suggested on the energetic grounds
\citep{Colgate00,Furlanetto01,Kronberg01}, without details of the exact physical
processes of how magnetic fields might be transported and amplified.  
Recently, using self-consistent cosmological MHD simulations, \citet{Xu09} showed that
magnetic fields from an AGN injected in a local region can be sufficient to magnetize
the whole cluster to the micro Gauss level through the action of cluster mergers 
and the ICM turbulence. This result, which is consistent with semi-analytic
models 
of the evolution of magnetic fields and turbulence \citep{Subramanian06}, shows that the small-scale 
turbulent dynamo \citep{Boldyrev04,Brandenburg05} may operate in the ICM.

In this paper, we investigate the influence of injection energy and redshift on the evolution of 
cluster magnetic fields and their properties at the present epoch. In a following
paper, we will study the magnetic field evolution of galaxy clusters with different masses
and at different dynamical stages.
Here, we present the properties of magnetic fields in the ICM
and the distribution of synthetic Faraday rotation measurement.
We also discuss the importance of high numerical spatial resolution in a large volume to follow the magnetic fields correctly
in cluster formation simulations.
The setup of our simulations is described in
Section \ref{sec:model} and results are presented in
Section \ref{sec:result}. Our results are summarized in Section 4.

\section{Basic Model and Simulations}
\label{sec:model}

Cosmological MHD simulations were performed using
the newly developed ENZO+MHD code \citet{Collins09}, which is an Eulerian cosmological
MHD code with adaptive mesh refinement (AMR). The initial conditions of our simulations
are generated from an \citet{Eisenstein99} power spectrum. The simulations use a $\Lambda$CDM model with parameters $h=0.73$,
$\Omega_{m}=0.27$, $\Omega_{b}=0.044$, $\Omega_{\Lambda}=0.73$, and
$\sigma_{8}=0.77$. These parameters are different from those used in previous 
simulations \citep{Xu09} and are close to the values from the recent WMAP observations \citep{Spergel07}.
The most important difference between simulations presented here and the previous one is the baryon to dark matter ratio, 
which is 14.76$\%$ (compared to 8.20$\%$ in \citet{Xu09}).
It is encouraging that the evolution of the magnetic field in this work
is similar to the evolution found in \citet{Xu09}, given the differences in
cluster properties and $\Omega_b$. The simulation volume is $351$ Mpc on a side, and it uses a $128^3$
root grid and $2$ level nested static grids in the Lagrangian region
where the cluster forms. This gives an effective root grid resolution
of $512^3$ cells ($\sim$ 0.69 Mpc) and dark matter particles of mass resolution of $1.07
\times 10^{10}M_{\odot}$. As the simulation proceeds, $8$ levels of refinement are allowed beyond 
the root grid, for a maximum spatial resolution of $11.2$ kpc.

The energy of an AGN is initially injected into the ICM in the form of magnetic fields (see description in \citet{Xu08a})
over the most massive halo locally. 
Part of the magnetic energy, however, is quickly converted to other energies (kinetic motion and thermal energy) during the 
formation of radio jets and lobes in the ICM \citep{Li06, Xu08a}.
Because it is currently not possible to resolve both the galaxy cluster and the AGN environment 
simultaneously, we have adopted an approach that mimics the possible magnetic energy injection 
by an AGN \citep{Li06}. The size of the injection region and the associated field strength 
are not realistic when compared to the real AGN jets, but on global scales ($>$ tens of kpc), 
the previous studies by \citet{Nakamura06} and \citet{Xu08a} showed that this 
approach can reproduce the observed X-ray bubbles and shock fronts \citep[e.g.][]{McNamara05,Nulsen05}.

The simulations were evolved from redshift $z=30$ to $z=0$
without radiative cooling, star formation, or feedback (so-called adiabatic). In this study, we ``turn on'' the AGN with magnetic fields at different redshifts of 
$z=3$, $z=2$, $z=1$ and $z=0.5$, centered at the most massive halo in the proto-cluster at the injection time.
We have two sets of simulations. First, we have six simulations of the AGN activated at z=3 with different magnetic energies.
The reason why our study concentrates on z=3 is that the comoving quasar number density peaks at z $\sim$ 2.5 \citep[e.g.][]{Fan01}.
 We label simulations of this set as runs A to F, from the most to the least injected energy.
The second set of simulations inject magnetic energy similar to run C, but with the injection
redshift varied. 
Simulations with injection at z=2, z=1, z=0.5 are labeled as z2, z1 and z05.
All the parameters of the injection model \citep{Li06} except the strength of the magnetic fields are kept the same for all the simulations.
The magnetic energy is injected inside 40 kpc (proper not comoving) and 
the injection lasts for 36.7 Myr.
The properties of the halo at the AGN injection and the initial and final magnetic energies are summarized in Table \ref{table:halos}.

\begin{table}
\caption{Halo Properties and Magnetic energy of simulations.}
\begin{center}
\label{table:halos}
\begin{tabular}{|c|c|c|c|c|c|c|c|}
\hline
\hline
  &   & \multicolumn{4}{c|}{Halo Properties} & \multicolumn{2}{c|}{Magnetic Energy}  \\
Runs & Injection z &  $r_{200}(Mpc)$ &  $M_{vir}(M_\odot)$ & $M_{gas}(M_\odot)$ & $M_{DM}(M_\odot)$ & Injected (ergs)  & z=0 (ergs) \\
\hline
A     & 3   & 0.189  & 1.50e13 & 2.14e12 & 1.28e13   & 1.94e60 &  1.43e61 \\
B     & 3   & \multicolumn{4}{c|}{same as above}  &   6.17e59  &  9.55e60  \\
C     & 3   & \multicolumn{4}{c|}{}               &   1.85e59  &  6.48e60   \\
D     & 3   & \multicolumn{4}{c|}{}              &     4.98e58 &  7.25e60  \\
E     & 3   & \multicolumn{4}{c|}{}               &   1.27e58  &  5.13e60   \\
F     & 3   & \multicolumn{4}{c|}{}               &   3.20e57  &  2.86e60   \\
z2     &  2    & 0.386  & 5.71e13 & 7.33e12 & 4.98e13  &  1.46e59  & 5.83e60 \\
z1     &  1    & 1.11   & 4.89e14 & 6.77e13 & 4.21e14  & 1.47e59 &   9.47e59 \\
z05    &  0.5  & 1.53   & 7.28e14 & 1.03e14 & 6.24e14  & 1.68e59 &  6.45e59 \\
\hline
\end{tabular}
\end{center}
\end{table}

AMR is applied only in a region of (50 Mpc)$^3$ where the galaxy cluster forms.
During the course of cluster formation, before the magnetic fields are injected, the refinement is controlled 
by baryon and dark matter overdensity, which is similar to other cluster formation simulations \citep[e.g.][]{Motl04,Nagai07}. 
After the magnetic field injection, in addition to refinement by overdensity,
all the regions where the magnetic field strength is higher than 
$5 \times 10^{-8}$ G are refined to the highest level. 
Using this refinement method, the volume that is refined to the highest level contains more than $99\%$
of the total magnetic energy \citep{Xu09b}.  For run A, the simulation is 
equivalent to $\sim$ 600$^3$ uniform grid ($\sim$ 600$^3$ zones at the highest level) MHD runs in the cluster region with full cosmology. 
For cases with lower injected magnetic energy, the highly refined region will be smaller (e.g. 
the volume with the highest refinement of run F is about half of that of run A). The data analysis in this paper is performed 
using yt\footnote{http://yt.enzotools.org} for Enzo \citep{Turk08}.

\section{Results}
\label{sec:result}

\subsection{Formation History of the Galaxy Cluster}
\label{sec:clusterformation}

\begin{figure}
\begin{center}
\epsfig{file=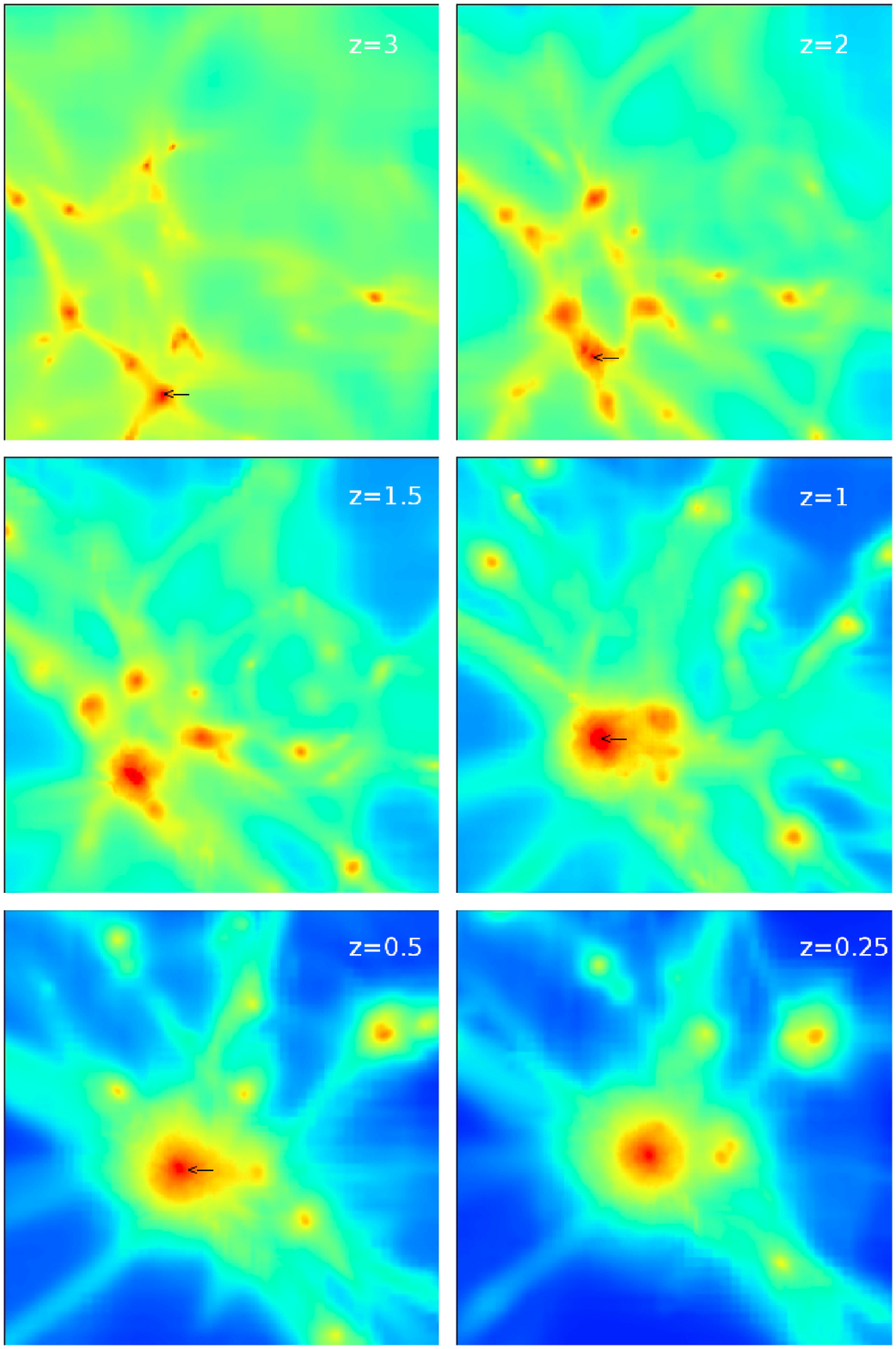,height=0.8\textheight}
\end{center}
\caption{Snap shots of the projected baryon density for different epochs of the 
  formation history of run A. Each image covers a region of
  $16$ Mpc $\times$ $16$ Mpc (comoving) at the simulation domain center. The projected results
  are obtained by integrating $16$ Mpc (comoving) centered at
  the simulation center along lines perpendicular to the observed plane. The six columns are marked
  with the respective redshift as $z$ = 3, 2, 1.5, 1, 0.5, 0.25.  The color range is
  from $1.0 \times 10^{-5}$ to  $1.0 \times 10^{-1}$
  g cm$^{-2}$ for the baryon density.
\label{fig:density_caseA}}
\end{figure}

Before we discuss the properties of magnetic fields in the ICM, we present 
the history of the galaxy cluster formation by hierarchical mergers. The merger history
determines the properties of the ICM dynamics and, in turn, the evolution of magnetic
fields. In Fig. \ref{fig:density_caseA}, we show the projected gas density at different 
redshifts corresponding to different stages of cluster formation for run A. The halos 
with AGNs injected at different redshifts are indicated by an arrow. The arrow in the z=3 panel applies
to runs A to F, while the arrows in the z=2, 1, 0.5 panels apply to runs z2, z1,
z05, respectively. As we will show later, the magnetic field injection
does not change the halo merger properties, so we can use the same projection
to indicate the AGN host halo for all simulations.
Each figure is a projection of baryon density along the y direction in a box of $(16~Mpc)^3$ comoving 
centered at the center of the simulation domain. 

\begin{figure}
\begin{center}
\epsfig{file=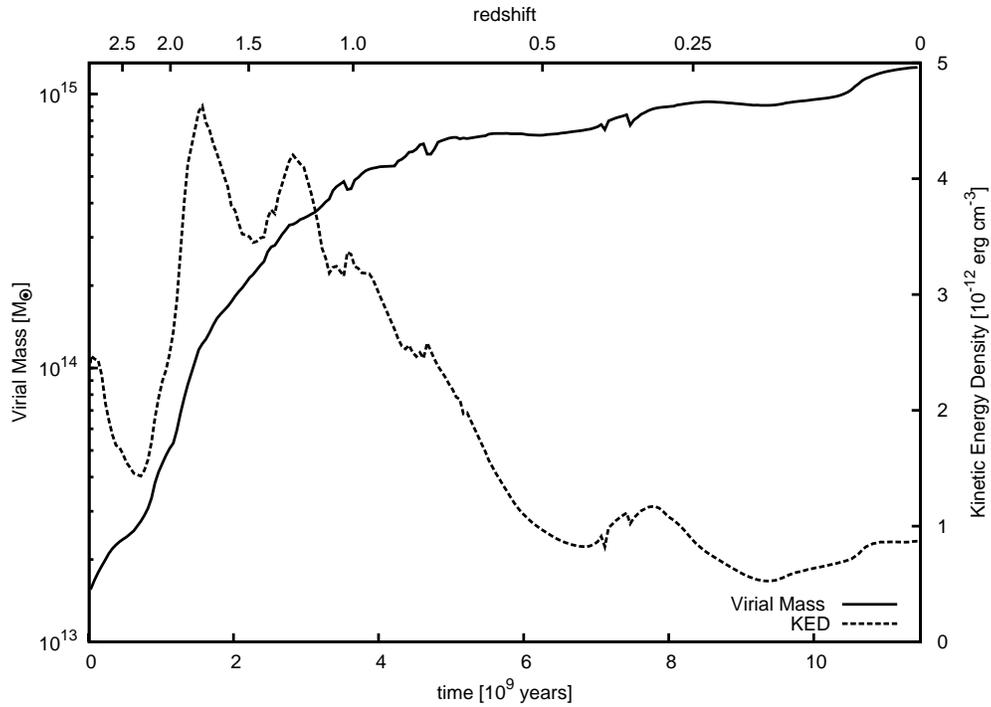,width=0.8\textwidth}
\end{center}
\caption{Virial mass and averaged kinetic energy density inside the virial radius as 
function of time of run A. The mass of the galaxy cluster increases rapidly
 before 5 Gyr due to many mergers, so the kinetic energy density remains high. After z=1, 
both the mass growth rate and the kinetic energy density drop because the 
merging halos are relatively small compared to the whole cluster.
\label{fig:Virialmass}}
\end{figure}

Fig. \ref{fig:Virialmass} 
plot the evolution of mass and averaged kinetic energy density inside the virial radius ($r_{200}$) 
of the cluster. This shows the mass growth of the cluster
and the kinetic energy available to amplify the magnetic fields. When we compute the
kinetic energy, we substract the averaged velocity of the whole cluster (the bulk motion
of the cluster that cannot be used to amplify the magnetic fields). 

Galaxy clusters are formed by hierarchical mergers as described by \citet{Motl04}. During the period 
we studied from z=3 to z=0, there are two different stages. From z=3 to z=1, the cluster is still small 
and increases in mass and size rapidly by frequent mergers with halos of similar sizes.  
From z=3 to 1, the mass of the cluster increases from 
1.5 $\times$ 10$^{13}$ to 5 $\times$ 10$^{14}$ $M_\odot$. 
Since the mass from major mergers is added to the cluster slowly (over a few 100 Myrs), the virial mass inceases quite smoothly.
Due to frequent mergers, 
the cluster is active and the kinetic energy density is relatively high.
The peaks of the evolution of kinetic energy density show the impact of major mergers. 
By z=1, the major part of cluster has been formed. Mergers have become rare and the merging 
halos are relatively small compared to the central cluster, producing smaller disturbance on the ICM. 
Thus the mass grows at a much slower rate (a factor of 2.5 for the remaining 7 Gyr) and the cluster becomes 
relaxed, with a lower kinetic energy density that decreases
gradually to only about one quarter its value before z=1. 
The cluster has a final virial mass
of 1.25 $\times$ 10$^{15}$ $M_\odot$  with a virial radius of 2.16 Mpc.

\begin{figure}
\begin{center}
\epsfig{file=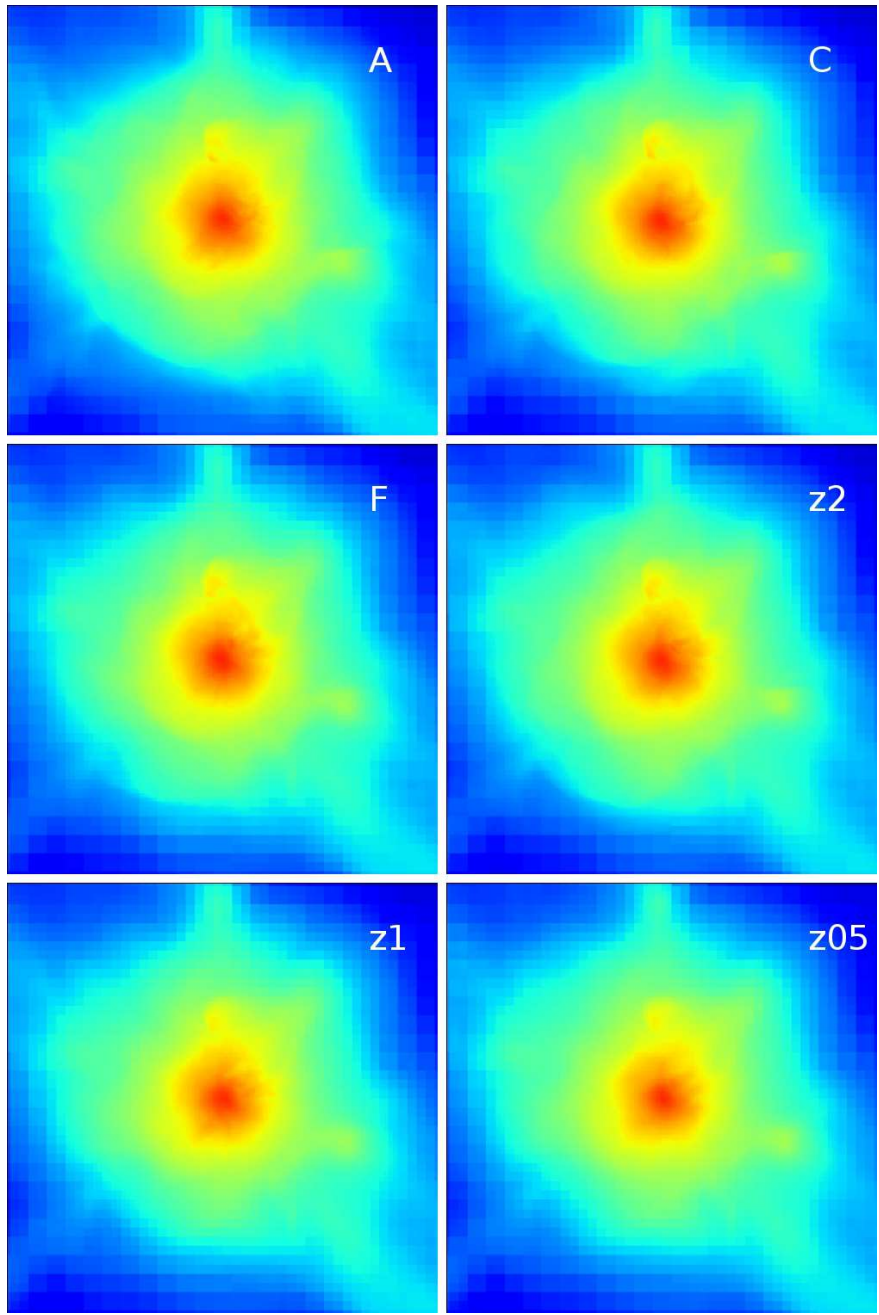,width=0.7\textwidth}
\end{center}
\caption{Projected baryon density for different injection levels of magnetic energy at $z=0$. 
  Each image covers a region of
  $8$ Mpc $\times$ $8$ Mpc. The projected results
  are obtained by integrating $8$ Mpc centered at
  the simulation center along lines perpendicular to the observed plane. The panels are marked
  with the respective cases.  The color range is
  from $1.0 \times 10^{-5}$ (blue) to  $1.0 \times 10^{-1}$ (red) g cm$^{-2}$.
\label{fig:densityz0}}
\end{figure}

\begin{figure}
\begin{center}
\epsfig{file=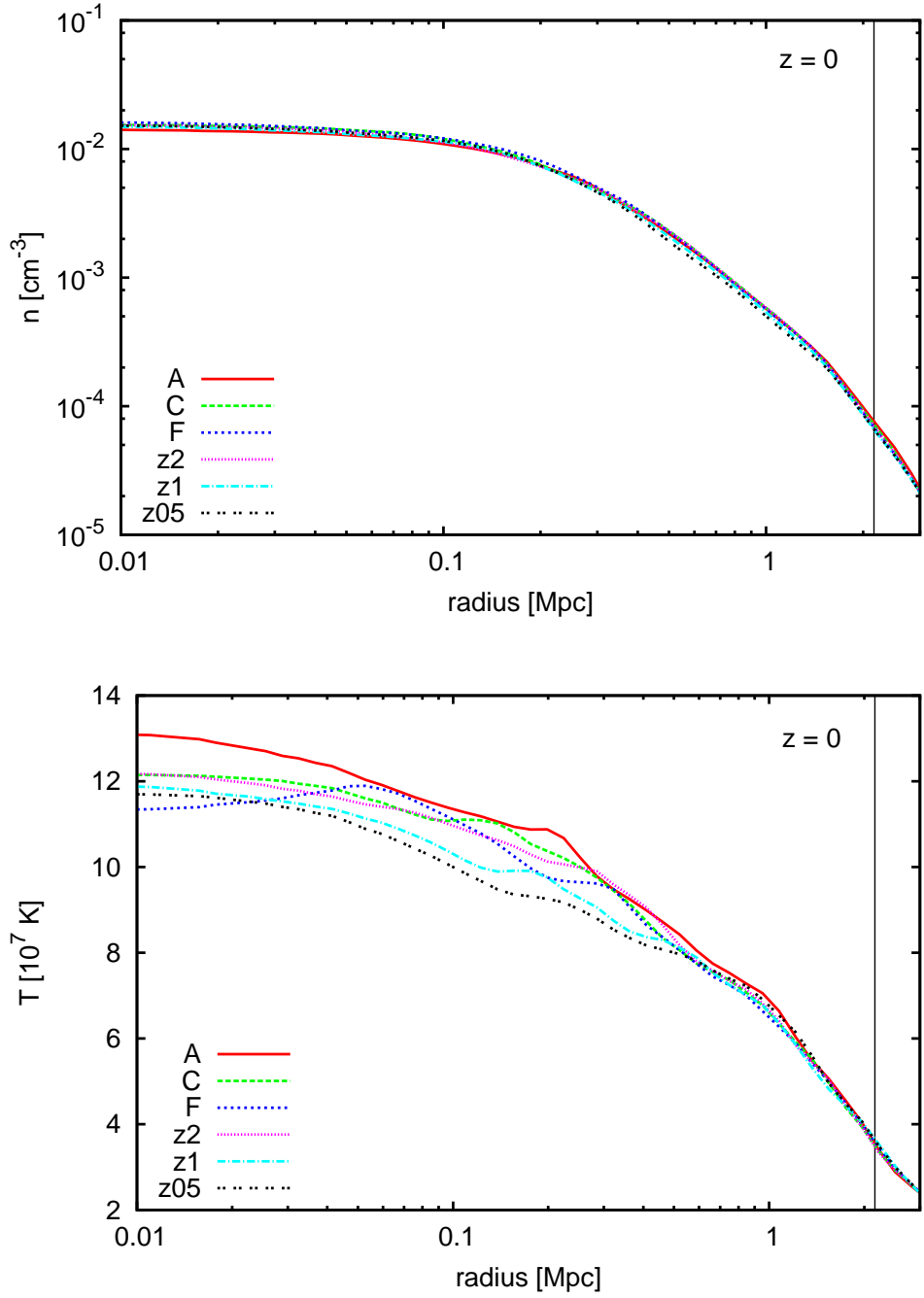,width=0.8\textwidth}
\end{center}
\caption{Spherically averaged radial profiles of the gas density and temperature at z=0. The magnetic fields
have small impact on the density and temerature profiles over the cluster central region. The solid vertical line shows the virial radius. 
\label{fig:profile_z0}}
\end{figure}

To show the influence of the magnetic fields 
on the cluster formation,  Fig. \ref{fig:densityz0} plots the projected gas density at
redshift z=0 for runs A, C, F, z2, z1, z05, in smaller projection boxes of $(8~Mpc)^3$, while Fig. 
\ref{fig:profile_z0} shows radial profiles of their baryon density and
temperature. Though the injected magnetic fields and the final field distributions 
are quite different for various simulations, the gas density and temperature distributions
are just moderately affected, mostly in the cluster central region.. The gas density increases and the temperature decreases with both
injection energy and injection time, but only at most 10\%. The magnetic energy in the clusters is much smaller than the kinetic and 
thermal energy (see \citet{Xu09} and Fig. \ref{fig:Beta_volume_histogram} 
in Sec. \ref{sec:profile}) throughout most of the cluster. 
So they are dynamically unimportant during the cluster formation. The impact of magnetic 
fields is more visible in the inner 400 kpc but these don't change the properties of the cluster as a whole significantly.


\subsection{Magnetic Field Distribution and Magnetic Energy Evolution}

\subsubsection{Magnetic Field Spatial Evolution}

\begin{figure}
\begin{center}
\epsfig{file=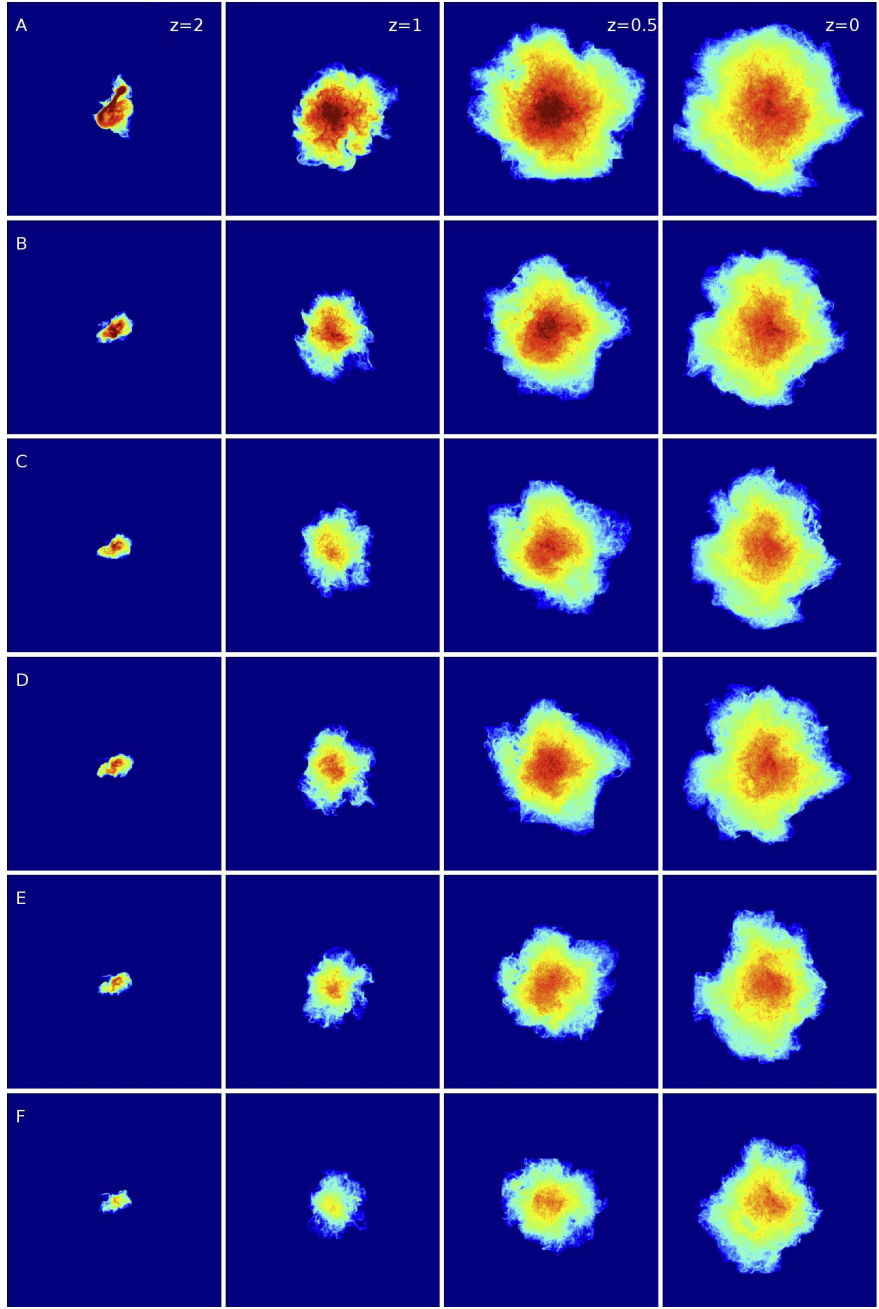,width=0.7\textwidth} 
\end{center}
\caption{Projected magnetic energy density of runs A to F at different redshifts. Each image is 8 Mpc 
 on a side and centered at the cluster center.
The projected results are obtained by integrating $8.0$ Mpc centered at
  the cluster center along lines perpendicular to the observed plane. The color range is
  from $1.0 \times 10^{8}$ (blue) to  $1.0 \times 10^{12}$ (red)
  ergs cm$^{-2}$.
\label{fig:med_z3}}
\end{figure}

\begin{figure}
\begin{center}
\epsfig{file=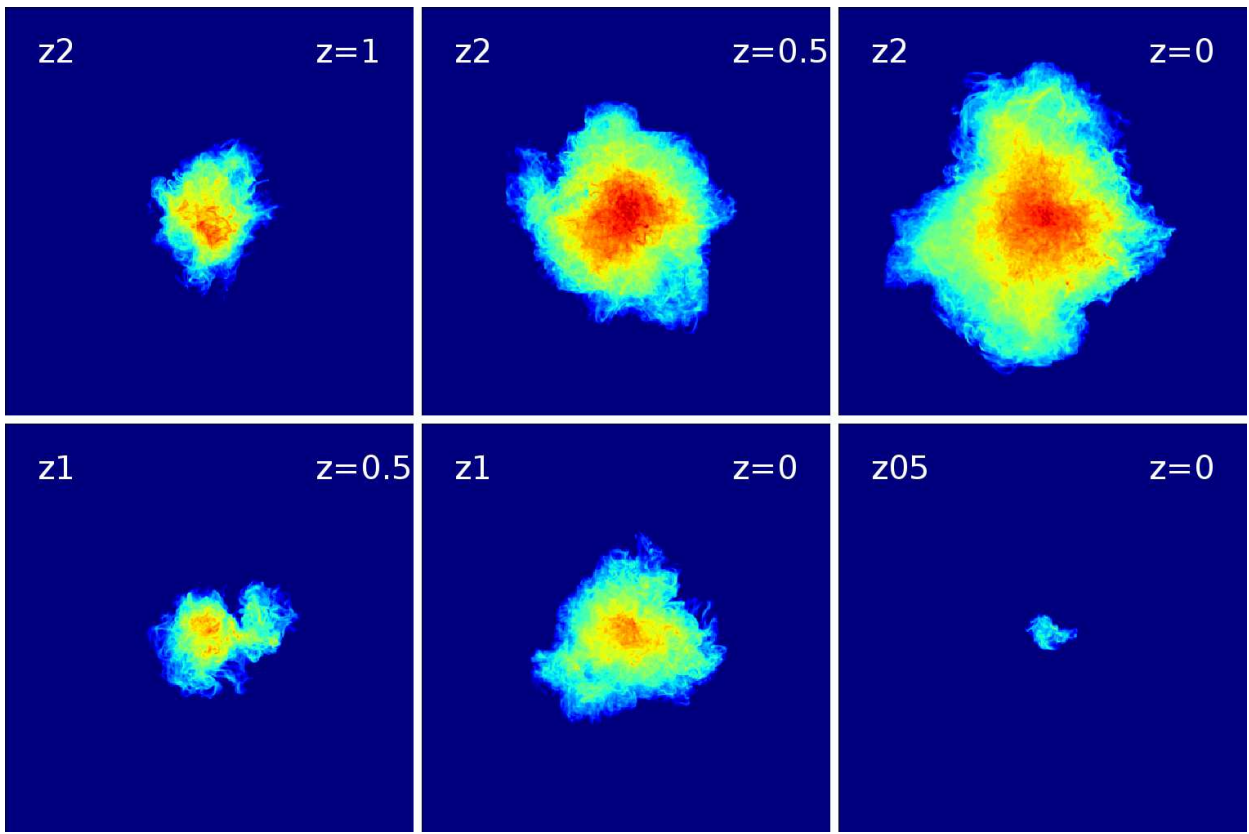,width=\textwidth} 
\end{center}
\caption{Similar to Fig. \ref{fig:med_z3}, except for runs z2, z1 and z05. 
\label{fig:med_z2}}
\end{figure}

To illustrate the spatial evolution of the magnetic fields, 
we present the projections of magnetic energy density ($B^2/8\pi$) at different
redshifts in Fig. \ref{fig:med_z3} for cases with AGN ``turned on'' at z=3 and in
Fig. \ref{fig:med_z2} for all other cases. We set the center 
of the view at the cluster center at the observed time, since
the magnetic fields  move with
the cluster as shown in Fig. \ref{fig:density_caseA}. With the strong local injections, 
the magnetic fields first expand by their own pressure to form bubbles. The mergers and
random motions of the ICM then destroy the bubbles and spread the magnetic fields away 
from the vicinity of the injection locations to the whole cluster. These processes were described in \citet{Xu09}. 

For cases with an AGN injected at z=3, the magnetic fields basically evolve in a similar way 
and are all distributed throughout a large volume of the cluster at low redshifts. The magnetic fields are spread throughout
the cluster very quickly before z=0.5 when the cluster is more active. The strength of the initial magnetic fields
determines the early expansion of the fields, thus the final field distribution.
Magnetic fields in run A, with the strongest injected fields, expand the fastest
at early times
and are distributed in the largest volume at any stage during the cluster formation. After $z=0.5$, the 
magnetic field distribution in the ICM shows signs of saturation. Note that the cluster 
is continuously growing in mass and undergoing minor mergers (albeit slowly). Here we 
regard that magnetic fields have reached a "saturated" state when they have stopped 
growing (or have very slow growth) both in their total energy (see Fig. 7 in the next subsection) and 
spatial distribution.  The detailed physical processes
of saturation, however, deserve a more careful evaluation, which will be 
presented in a future publication. Though expanding with different rates at early time, magnetic field 
distributions of runs B, C, and D are quite similar at z=0.  It seems that they
are all saturated at z=0, and this saturation level is determined more
by the ICM properties than the initial magnetic field strength. For runs E and F, the magnetized regions are significantly 
smaller than the stronger injection energy cases, and the
strength of projected magnetic fields is  weaker.
 Their magnetic fields have more difficulty
 expanding in the ICM at the early time due to their weak initial 
magnetic pressure.  This leads to a much smaller
cross section for the mergers, and consequently less chance of being brought to the other parts of the
cluster and to be amplified. They continue to expand to 
a much larger volume from z=0.5 to z=0, and don't clearly show
indication of being saturated.
 
For AGN injected at z=2, which injects magnetic fields about 1 Gyr later, the magnetic fields are also spread throughout the whole cluster. Though the 
initial magnetic fields are injected at a different places, and
with larger energy, the field distribution at z=0 is very similar to runs C and E.
For the run with injection at z=1, magnetic fields still spread out of the
cluster center, but are distributed in a much smaller area. 
This is because the injected magnetic fields miss the most active period of the cluster mergers. For run z05, 
though the magnetic bubbles are still destroyed, the
magnetic fields are  confined in the 
central region of the cluster. The cluster at z=0.5 is already quite large,  so it is difficult for the 
mergering halos to bring the magnetic fields far away from the center. For the case when the AGN 
injection is much later at z=0.05 \citep{Xu08a}, the 
magnetic fields could survive in a bubble morphology, which 
resembles observations of X-ray cavities in clusters.

\subsubsection{Magnetic Energy Evolution}
\label{sec:energyevolution}

\begin{figure}
\begin{center}
\epsfig{file=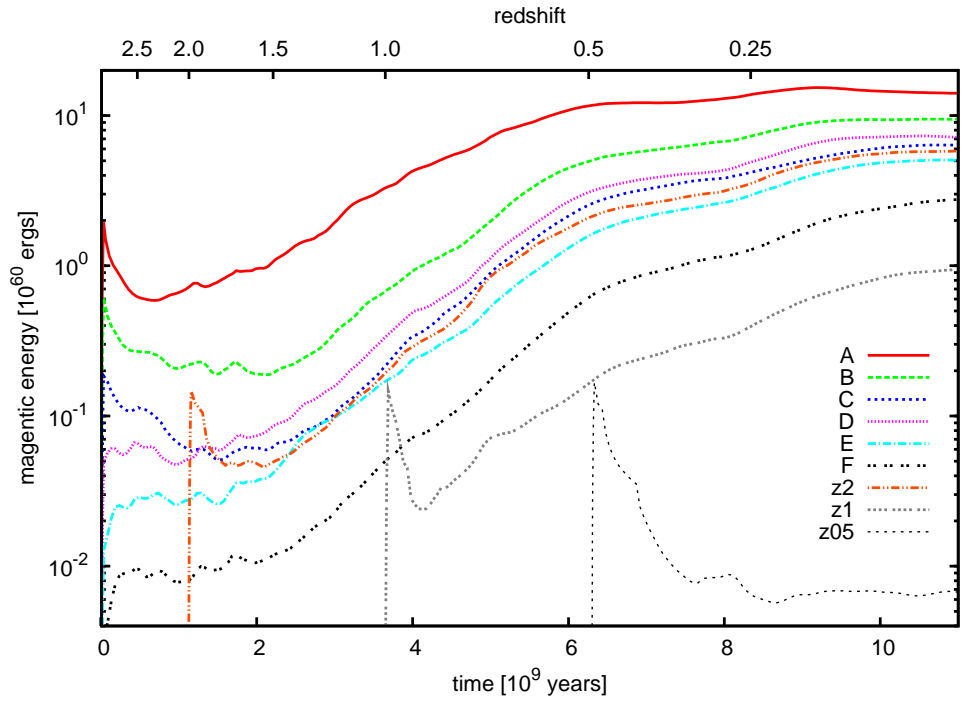,width=0.8\textwidth} 
\end{center}
\caption{Evolution of total magnetic energy of 
  all simulations. 
\label{fig:energy_z3}}
\end{figure}

Since we have only one source of magnetic fields in each simulation, it is straight forward to track the history of magnetic energy 
amplification in the cluster. The evolution of the total magnetic energy of  all the simulations is shown in Fig. \ref{fig:energy_z3}.

The magnetic energy generally decreases rapidly initially for a few hundred
million years because of the rapid expansion of the magnetic structure. For AGN injected before the main mergers (runs A-F, z2 and z1), the
magnetic energy gradually increases due to the combined effects of major mergers and the ICM turbulence until saturation occurs or the simulation ends.

At z = 0, the total magnetic energy is (14, 9.6, 6.5, 7.3,  
5.1, 2.9) $\times$ 10$^{60}$ ergs, for runs A-F
with a gain of about 10, 15, 40, 150, 400, and 900  times, respectively. It is interesting 
that the magnetic energy of run F, which has an injected energy of just 3 $\times$ 10$^{57}$ ergs, or one-thousandth of 
the magnetic energy injected in run A, is one quarter the energy of run A at the final stage. 
So a small amount of magnetic fields from AGNs can still be amplified to 
magnetize the whole cluster to the observed level, provided that the ICM turbulence is strong.

For the later injection cases, the situations can be quite different.
For z05, the magnetic energy never increases, and drops to 7 $\times$ 10$^{57}$ ergs at z=0. As shown in the magnetic field distribution, 
since the magnetic fields are put into the system too late, there
is not enough time, or enough mergers, to distribute the magnetic fields to large volumes.  For the other two runs, z2 and z1, the magnetic fields still have enough time 
to expand and get amplified in the ICM turbulence. Their magnetic energy increases in a very similar way to cases of the injections at z=3. 
Run z1 has a final magnetic energy of 10$^{60}$ ergs, and the magnetic energy
evolution of z2  is almost identical to the run C after 3 Gyr, and seems saturated.

\subsubsection{Radial Profiles of Magnetic Fields}
\label{sec:profile}

\begin{figure}
\begin{center}
\epsfig{file=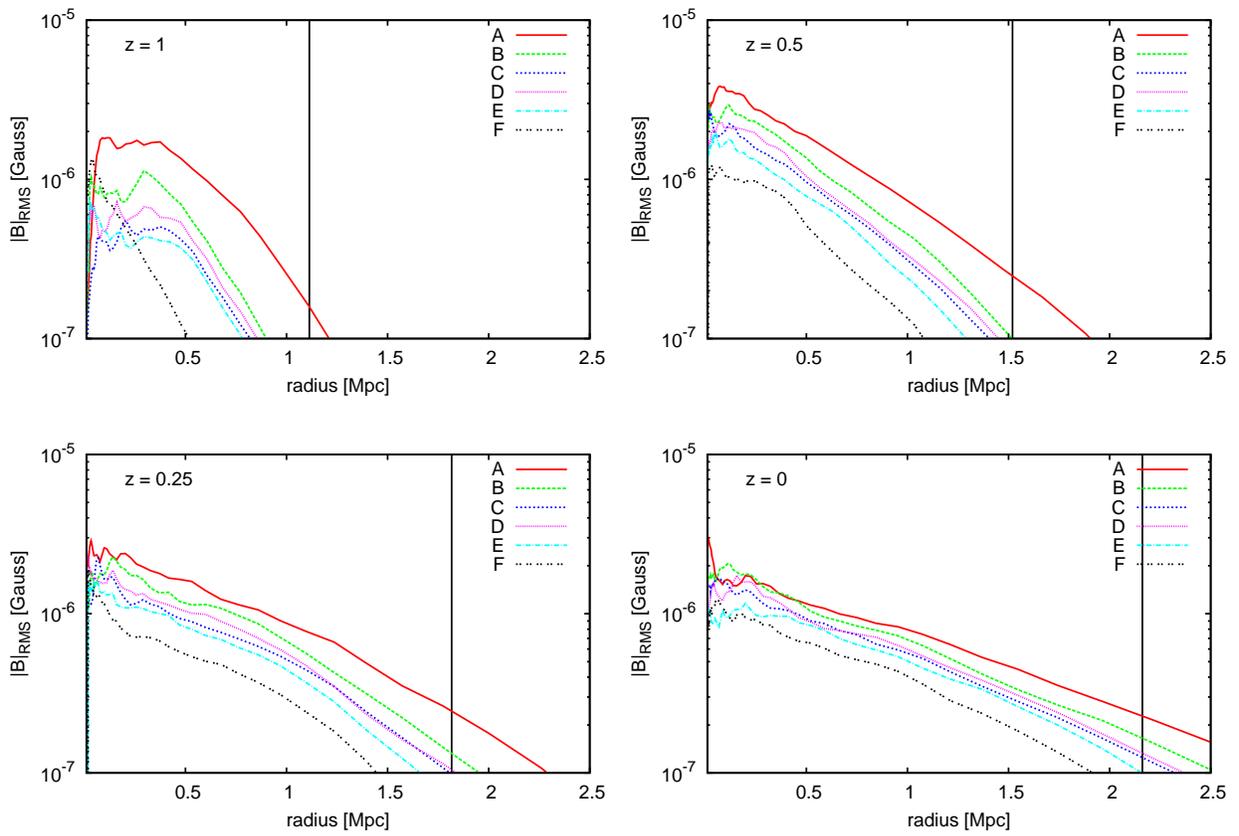,width=\textwidth} 
\end{center}
\caption{The spherically averaged radial profile of RMS magnetic field strength at different epochs of the
  cluster formation. The radius is measured in the proper distance 
Virial radii are marked by vertical lines.
\label{fig:Bprofile}}
\end{figure}

In Fig. \ref{fig:Bprofile}, we present the spherically averaged radial
profiles of RMS magnetic field strength at low redshifts of runs A to F. 
By z=1, the strong initial magnetic field of run A allow it to reach the $\mu$G
level, and fill the virial radius of the cluster.  By z=0.5, all runs A-F have
field strengths of $>1 \mu\rm{G}$ at the cluster center.  By z=0, only the
run with the weakest initial field strength, run F, has not filled the virial
radius.

From z=1 to z=0, the magnetic field gradually saturates with the ICM motions from the center of the cluster to the virial radius.  
Magnetic field strength in the cluster center is maintained at the micro Gauss
level, while at larger radii, the field strength is continues to be amplified.
There is a clear trend that the slope of the radial profile decreases with time.
In run A, the magnetic field strength at the virial radius remains at a relatively
constant value of 0.2-0.5 $\mu$G from z=0.5 to z=0.  The magnetic field strenth in 
the other cases continues to be amplified at the virial radius in this reshift
interval, and their field strength profiles in these runs approach the
distribution in run A.
The run with the weakest injected magnetic field strength, run F, has its field
amplified fastest, and over the widest range of radii. 
 It seems that the profiles of magnetic fields are decided by
the ICM motions, rather than the initial magnetic field strength, provided there
is enough time for the magnetic field to be mixed throughout the cluster and
amplified by the turbulence.

\begin{figure}
\begin{center}
\epsfig{file=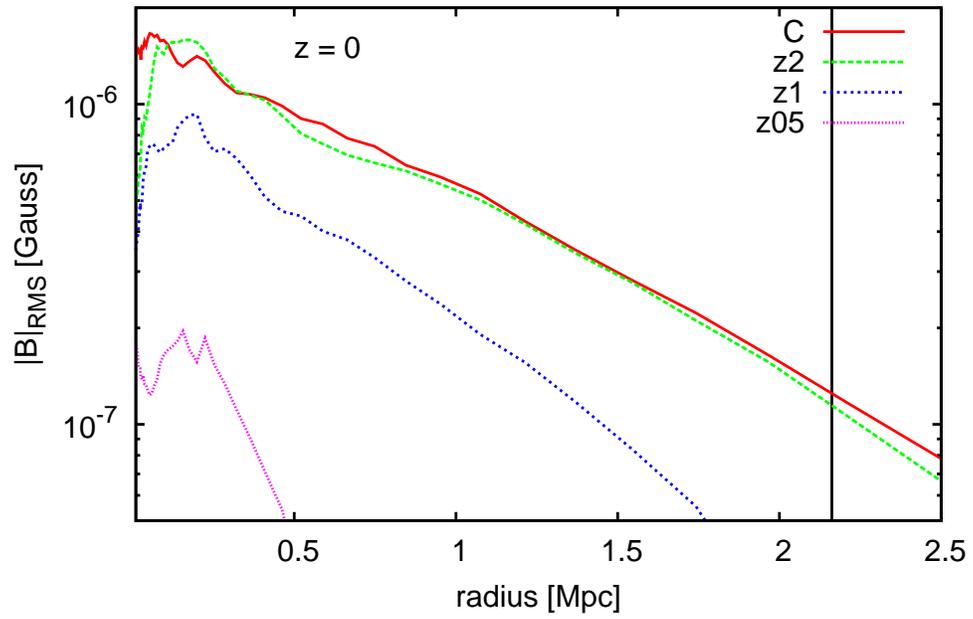,width=0.8\textwidth} 
\end{center}
\caption{The spherically averaged radial profile of RMS magnetic field strength at z=0 of runs C, z2, z1, z05.
\label{fig:B_profile_z0}}
\end{figure}

Fig. \ref{fig:B_profile_z0} shows the radial profiles of magnetic field strength
at z=0 for simulations with magnetic field injected after z=3. 
The magnetic fields of z2 are also higher than 1 $\mu$G and have almost the same profile as run C, except in the central 100 kpc. 
For run z1,  the magnetic
field strength is well distributed through most of the cluster and gets
amplified, while the final field strength is at the micro Gauss level, and spread 
out to more than 1.5 Mpc away from the center. But since there is less time for
the field to be amplified in the outer part of
the cluster, the radial profile is steeper.  
The magnetic fields of z05 is much weaker and locally distributed. 
It shows that the magnetic field injected this late are not likely to be an
important sources of cluster magnetic fields.

At z = 0, the magnetic field strengths in all cases injected before z=1 are at
the micro Gauss level at the cluster center, and
extend to large radii. Magnetic field strength profiles in these cases tend to a
power law, with best fit slope of -0.6 between r = 0.2 to 2 Mpc. This is flatter
than magnetic profiles seen in some other 
simulations \citep{Dolag02, Donnert08}. One possible reason for the flatter
radial profiles in our simulations 
is that we maintain a high resolution in regions with lower density but
significant magnetic field strength.
We have also performed a simulations with a less agressive set of refinement
criteria, the usual refinement criteria of gas and dark matter overdensity.
This causes lower resolution in the outer regions, where the density is low but
the magnetic field is significant.
We see a faster drop in magnetic field strength with lower resolutions at large radius (details will be discussed in Sec. \ref{sec:comparision}).  
The difference in profiles could also be caused by the different initial magnetic fields used in our simulations (vs uniform initial fields, for example). 
For instance, the gas collapse plays a less important role in the magnetic field amplifications in our simulations, since there is no magnetic field initially in the collapsing gas.

\subsubsection{Magnetic Field Volume Distribution}

\begin{figure}
\begin{center}
\epsfig{file=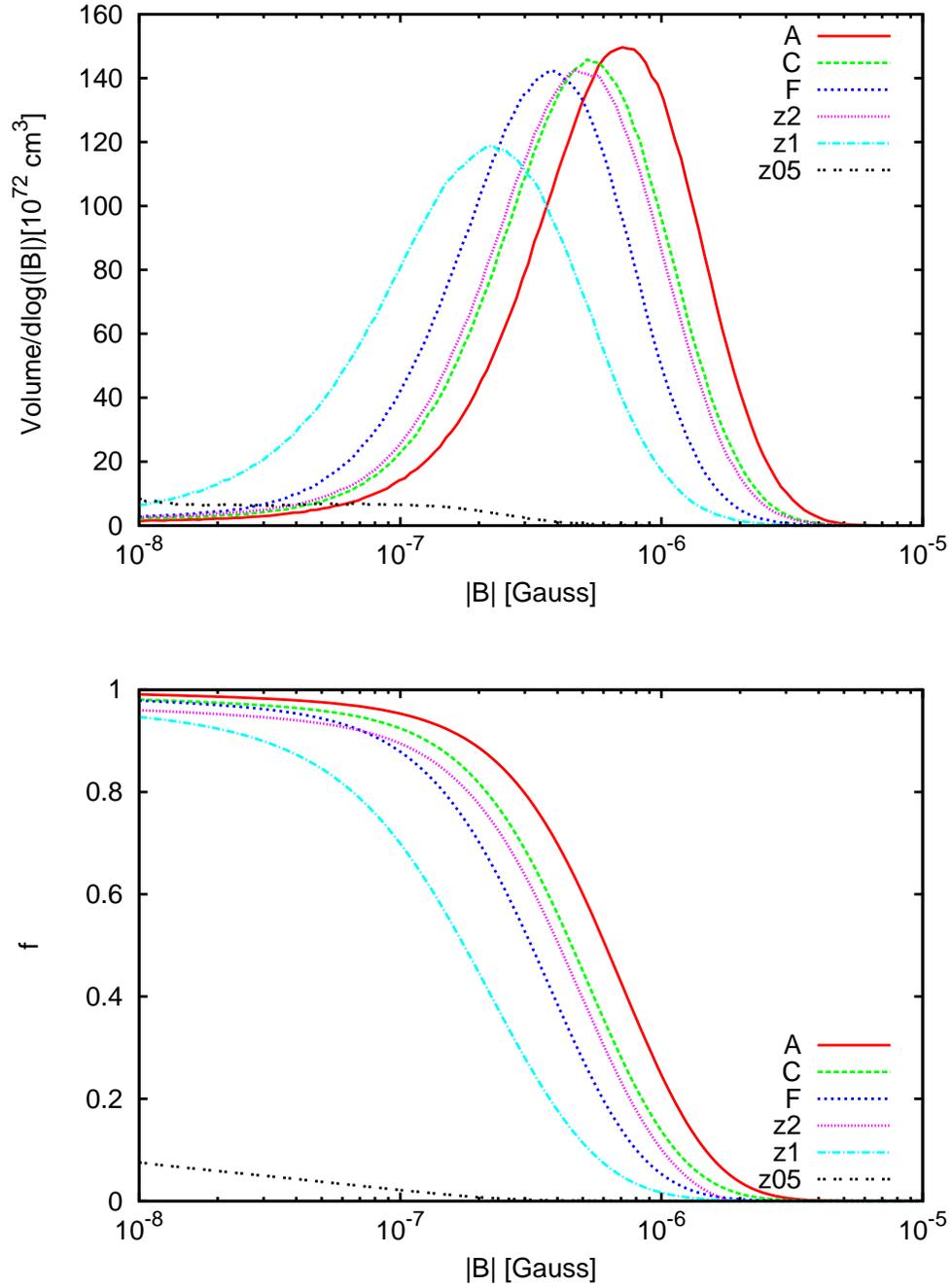,width=0.8\textwidth} 
\end{center}
\caption{Volume histogram (Top) and complementary cumulative volume histogram ($f(B) = \int_{B}^\infty P(B') dB'$) (Bottom) of magnetic field 
 strength of the central 1 Mpc sphere at z = 0.
\label{fig:MED_volume_histogram}}
\end{figure}

\begin{figure}
\begin{center}
\epsfig{file=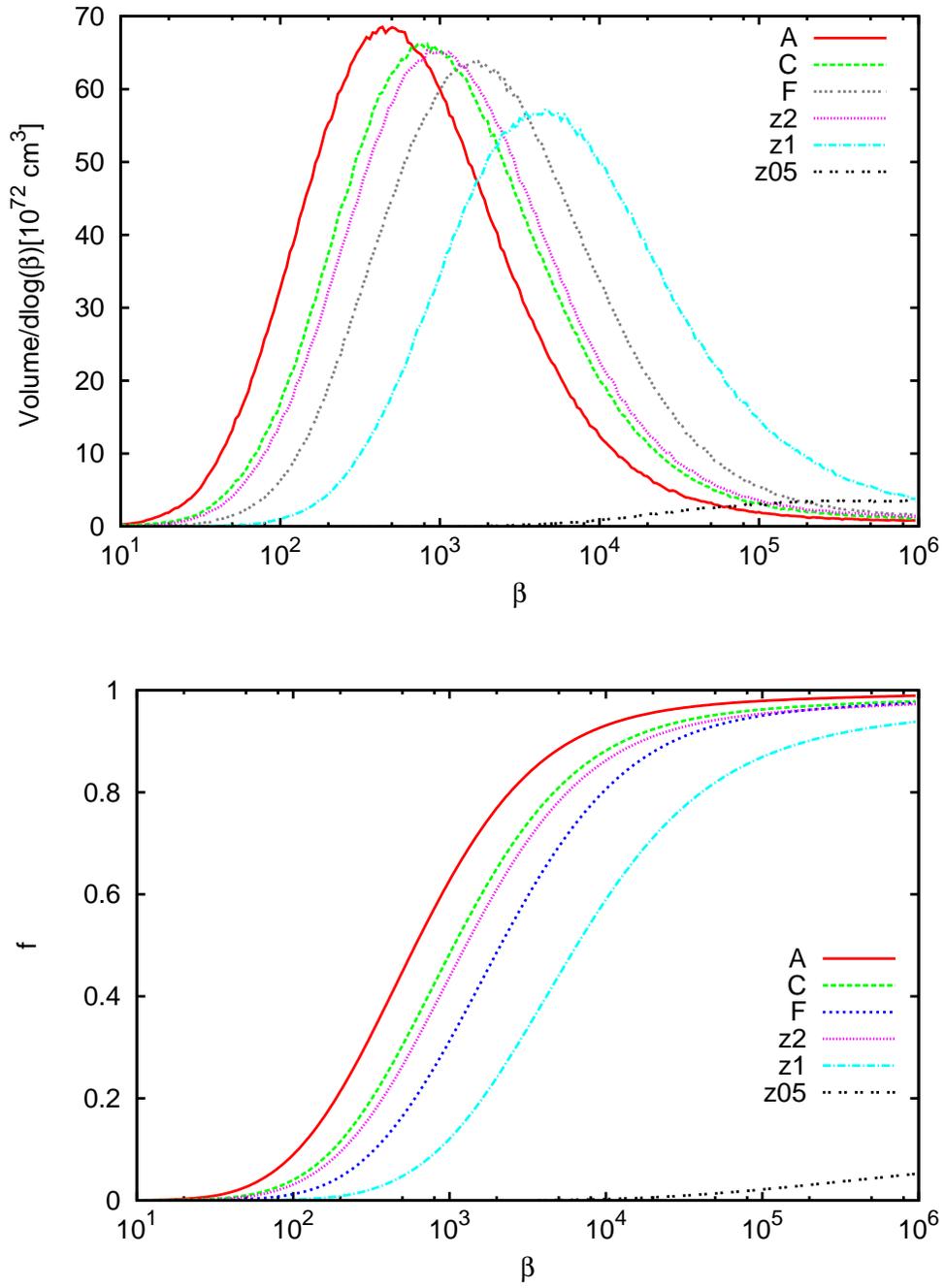,width=0.8\textwidth} 
\end{center}
\caption{Volume histogram (Top) and cumulative volume histogram (Bottom) of plasma $\beta$
  of the central 1 Mpc sphere at z = 0. 
\label{fig:Beta_volume_histogram}}
\end{figure}

In Fig. \ref{fig:MED_volume_histogram}, we plot the volume histograms of magnetic field strength of the central 1 Mpc sphere to demonstrate the 
spatial distribution of the magnetic fields.
Magnetic fields are mixed with the ICM plasma very well for all cases with injections before z=2.
More than 90\% of the ICM volume is populated with magnetic fields $\geq$ 0.1
$\mu$G. The peak of magnetic field strength distribution increases
with stronger or earlier magnetic field injections. On the other hand, magnetic fields of run z05 are very weak and fill less than 10\% of the volume.

Figure \ref{fig:Beta_volume_histogram} illustrates the volume histograms of the
plasma $\beta$ (=nk$_{B}$T/$\frac{B^2}{8\pi}$) at z=0 for the same volume. 
The thermal energy is always much larger than the magnetic energy. The peak of
the $\beta$ distribution is at $\sim$ 500 for run A.
The fact that $\beta >>1$ for these runs indicates that in most of the region, magnetic fields
are dynamically unimportant and follow the gas passively.

\begin{figure}
\begin{center}
\epsfig{file=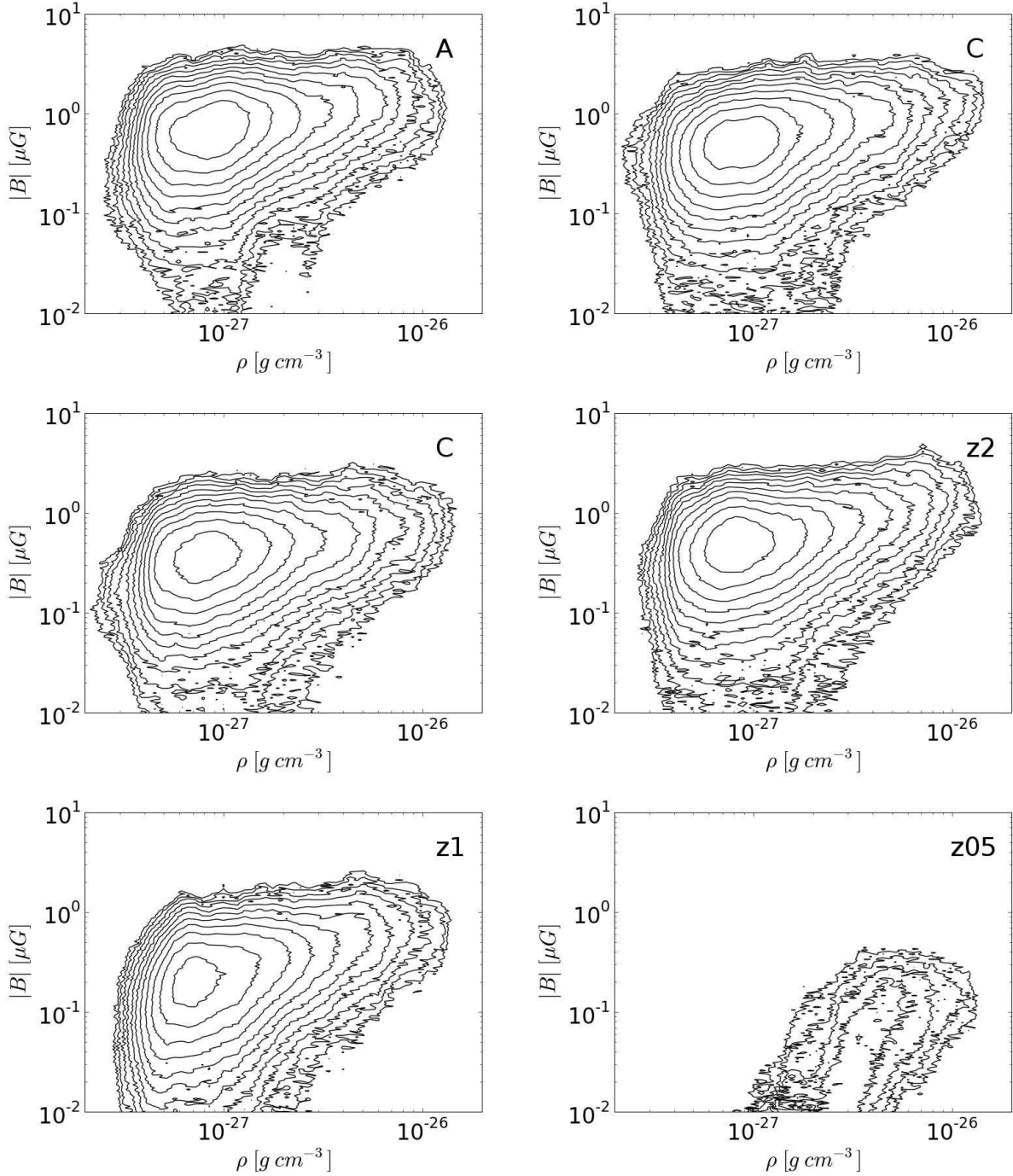,width=\textwidth} 
\end{center}
\caption{Volume weighted two-dimensional distributions of the magnetic field strength vs. baryon density of  
runs A, C, F, z2, z1, z05 inside the cluster core of 1 Mpc sphere at z = 0.
Contour lines are the volume of gas at that density and
magnetic field at 10$^k$ cm$^3$, where k =  69.0, 69.2, 69.4, ... 71.2 from outer to inner.
\label{fig:MED_density}}
\end{figure}

To show the correlation between the gas density and the strength of magnetic fields, we plot in Fig. \ref{fig:MED_density} 
the two-dimensional distribution of the magnetic field strength vs. the gas density at z=0 for runs A, C, F, z2, z1, and z05. 
There is no obvious correlation between the field strength and the gas density. The distribution is similar 
for the first five cases. Most of the magnetic field strength is between 0.1 to
a few micro Gauss, and are mixed with plasmas 
with a wide range of densities. This is because the magnetic fields are originally injected in a small volume 
and are randomly spread through the gas and amplified by the ICM turbulence. So
the field strength does not scale as 
$\rho ^{2/3}$, as in it would if the magnetic field was initially distributed in a large volume and mostly amplified by collapse. This property 
may be used to distinguish different 
magnetic field origins. The magnetic field strength in z05 is very weak and
mostly confined to the high density cluster core region. 

\subsection{Energy Power Spectrum and Small Scale Dynamo with MHD Turbulence}

\begin{figure}
\begin{center}
\epsfig{file=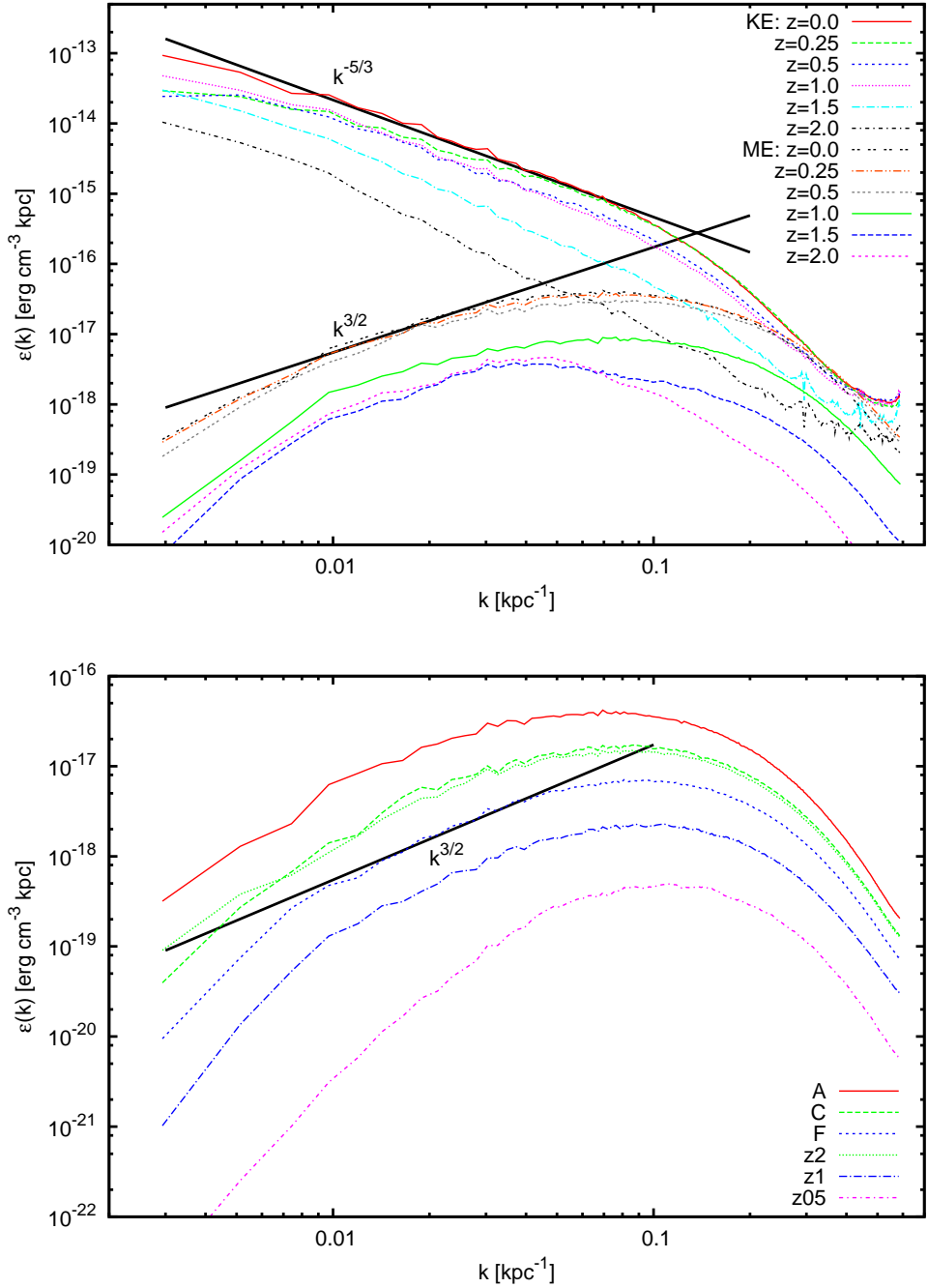,width=0.8\textwidth} 
\end{center}
 \caption{Top panel: Power spectra of the kinetic energy density and magnetic
   energy density of the ICM of run A at different redshifts. The ICM turbulence
   is represented by the Kolmogorov-like spectra in kinetic energy. The magnetic energy
   spectra follow the k$^{3/2}$ Kazantsev law in large scale.
   Bottom panel: Magnetic energy spectra of runs A, C, F, z2, z1, z05. 
   All sprecta except run z05 show a short k$^{3/2}$ range.
   \label{fig:power}}
\end{figure}

The top panel of Fig. \ref{fig:power} plots the power spectra of kinetic
and magnetic energy density of a box of 512$^3$ ($\sim$ (5.5 Mpc)$^3$
comoving) enclosing the cluster for run A at different redshifts. The
kinetic power spectra grow rapidly from z=2 to z=1 during the active
merger period of the cluster formation, then remain relatively constant  
level for the rest of the simulation. The kinetic power spectra show a
smooth power law similar to a Kolmogorov-like spectrum for incompressible turbulence
in the range k $\sim$ 0.01-0.1 kpc$^{-1}$. They are similar to
the spectra obtained from pure hydrodynamic simulations \citep{Vazza09}.
The spectrum steepens gradually for k $>$ 0.1 on small scales due to several
effects: the
damping of kinetic energy by the shocks \citep{Stone98};
the back reaction of magnetic fields, especially at low redshifts when
the magnetic energy density is close to the kinetic energy density at
those scales; and numerical diffusion. The magnetic energy power
spectra
at large scales follow the $k^{3/2}$ Kazantsev law.  At early times, when
the magnetic energy density is much smaller than the kinetic
energy density at all scales, the magnetic energy grows exponentially in all scales.
The power laws of both the kinetic and magnetic energy, and the
exponential growth of magnetic energy clearly show that the small scale
dynamo is operating in the ICM \citep[see][]{Brandenburg05}. The
characteristic timescale for magnetic energy growth is the eddy turnover
time at the dissipation scale (at k $\sim$ 0.1 to 0.2 kpc$^{-1}$ in this
simulation), which is $\sim$ 1 Gyr from z=2 to z=1, and increases
gradually to 3 Gyr at z=0. With the rapid growth from z=2 to z=0.5, the
magnetic energy density reaches equipartition with the kinetic energy
density at small scales. Then, the operation of small scale dynamo goes
into the second stage. During this stage, the magnetic energy grows
linearly with a rate proportional to the turbulence energy cascade rate
\citep{Schekochihin07} and the peak of magnetic energy moves to larger
scales \citep{Cho00}. The characteristic timescale is again the eddy
turnover time at the dissipation scale, which is $\sim$ 2 Gyr at z=0.5 and is
longer than 3 Gyr at z=0. The longer eddy turnover time is due to the
relaxation of the cluster. The linear growth of magnetic energy happens between
z=0.37 and z=0.16 at a rate of $\sim$ 1.5 $\times$ 10$^{60}$ ergs/Gyr.
After that, the magnetic energy stops growing and even decreases slowly.
We consider the magnetic fields saturated at this time, even though
the magnetic and kinetic energy have not reached equipartition globally.
Here, the saturation of magnetic fields happens when the operation of
small scale dynamo is in the linear growth stage and the relaxation of the
cluster makes this linear growth of magnetic energy very slow with the
long eddy turnover time.

The magnetic power spectra of six runs of the same volume at z=0 are
plotted in the bottom panel of Fig. \ref{fig:power}. We don't show the
kinetic power spectra since they are all almost the
same as that of run A. However, the magnetic power spectra for different runs
are at quite different levels. All the magnetic energy density power
spectra show a power law $k^{3/2}$ over a small range, except for run z05, so the magnetic
fields are indeed amplified by the small scale dynamo process in all these
runs. Magnetic fields in simulations with weaker initial magnetic fields
or later injection need more time to spread throughout the cluster, and
occupy relatively smaller volumes, especially for runs F and z1. The
evolution of magnetic energy shows that the magnetic fields of runs C and
z2 are
also in the second stage of small scale dynamo with the linear growth. The
magnetic energy density of these runs saturates at a subequipatition level
with respect to the kinetic energy density for
k $>$ 0.3 kpc$^{-1}$, which is likely because their magnetic fields fill
smaller volumes of the cluster. The magnetic energy of run F and z1
continues to exponentially grow for z $<$ 0.5 with a lower rate due to the
longer eddy turnover time and
does not clearly show whether or not the increase becomes linear at z
$\sim$ 0. For run z05, there are not enough big mergers to spread the
fields to large enough volume to catch the turbulence motions.
Even though there has been 6 Gyr since the AGN input, its power spectrum
shows no sign of the operation of the small scale dynamo.

Even though we have solved the ideal MHD equations in our simulations,
there is numerical diffusion that has allowed the turbulence to damp and
the magnetic fields to diffuse in the ICM. The rate of diffusion is
determined by the numerical Reynolds number and magnetic Reynolds number.
We estimated that these numbers in our simulations are on the order of a
few hundred. The real ICM could have Reynolds number is on the order of
100 or may be bigger (up to $\sim$1000) if magnetic fields are present
\citep{Sunyaev03}, which is close to our simulations.
But its magnetic Reynolds number is not well determined, especially in a
magnetized turbulent medium. The small scale dynamo theory and simulations
\citep{Haugen04} have shown that the dynamo will operate under our
simulation conditions. Most of the previous research of the small scale
dynamo is in incompressible turbulence limit. For the ICM, the compression
by shocks generated by mergers are very frequent and important, and they
could modify the turbulence properties. This in turn could affect the
operation of the small scale dynamo. Further study on the small scale
dynamo in compressible turbulence is needed.

\subsection{Faraday Rotation Measurement}

\begin{figure}
\begin{center}
\epsfig{file=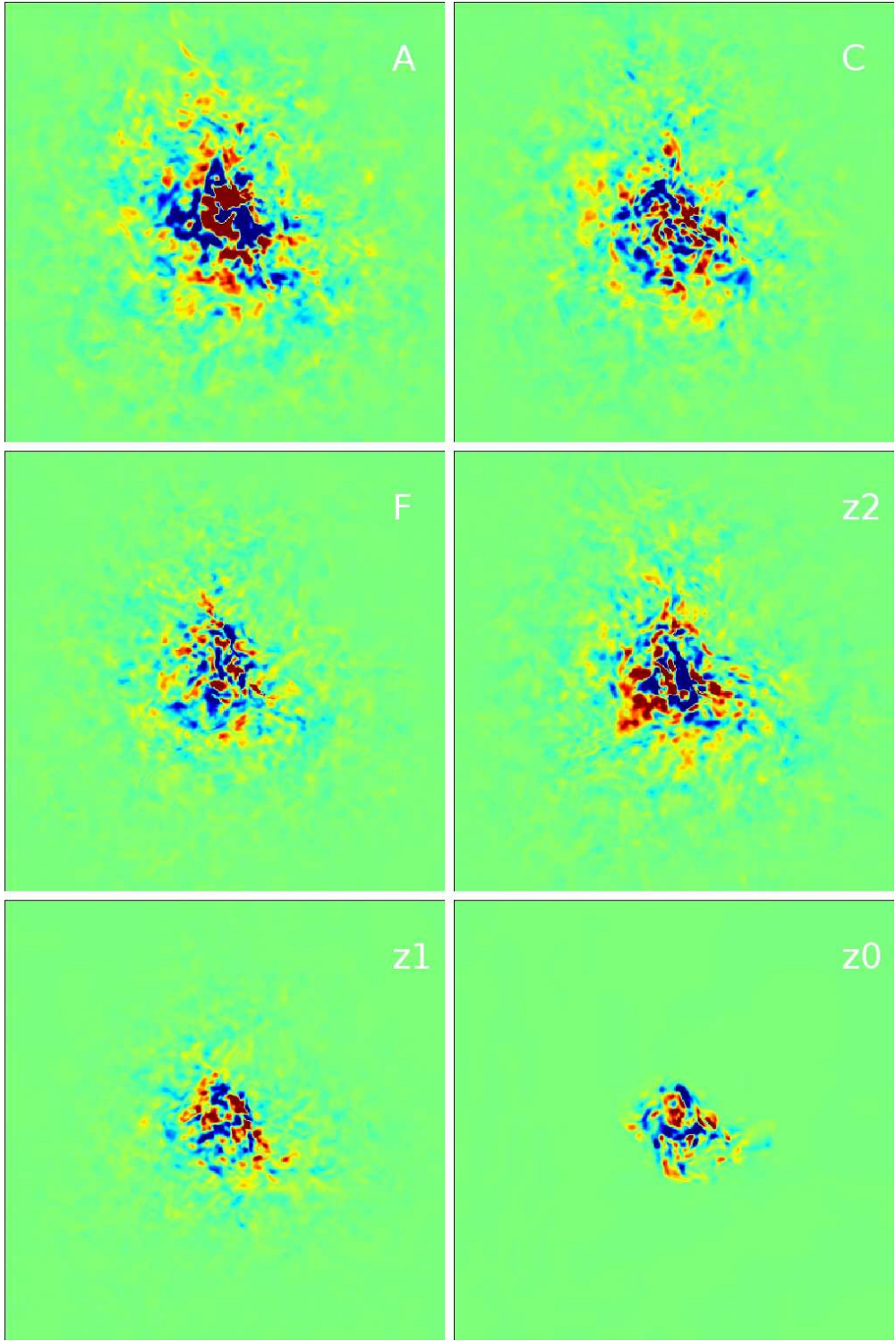,width=0.8\textwidth} 
\end{center}
\caption{Faraday rotation measurement of the ICM by integrating
  through the cluster along the y direction for different runs. It covers a region of $3$ Mpc  
  $\times$ $3$ Mpc at z = 0.  The color range shown is from $-500$
  (blue) to $500$ (red) rad m$^{-2}$.
  \label{fig:rm}}
\end{figure}

Another important observational description of magnetic fields in the ICM is the
FRM. The RM maps are used not only to determine the strength 
of magnetic fields but also the turbulent structure of them \citep{Ensslin06}. For a 
distant polarized source, the FRM, which is defined as ratio of change of polarization angle to
wavelength squared, in unit of rad m$^{-2}$, is \citep{Kronberg08}: 
\begin{equation} 
RM(z_s) = \frac{\delta \chi}{\lambda ^{2}} = 8.1 \times 10^5  \int_{0}^{z_s}
\frac{n_{e}(z) B_{||}(z)}{(1+z)^2} \frac{dl}{dz} dz,
\end{equation}
where $n_e$ is free electron number density in cm$^{-3}$, $B_{||}$ is the
line of sign component of magnetic fields in Gauss, and $dl/dz$ is the comoving 
path increment per unit redshift in parsec. We compute the synthetic 
FRM maps by integrating 8 Mpc along each axis.
The RM maps observed from different directions are quite similar, so we only show the results along the y axis here. 
Fig. \ref{fig:rm} shows the spatial distribution of FRM at z = 0 for six cases. 
The RM maps of runs with different initial field strengths or of different
injection times have similar features
but are different in details. Interestingly, the FRM map shows not only the small scale 
variations reminiscent of the ICM MHD turbulence, but also show long, narrow filaments, 
which may be relics of recent mergers. There are some high $|RM|$ regions at the outer
part of the cluster, which are consistent with the $>$ 100 rad m$^{-2}$ RM observed far from the cluster center in 
Abell 2255 \citep{Govoni06} and Coma cluster \citep{Bonafede10}. For run z05,
the FRM not only covers the smallest area, but has much 
less small-scale structure, which is because the magnetic fields has had little amplification on small scales. 

\begin{figure}
\begin{center}
\epsfig{file=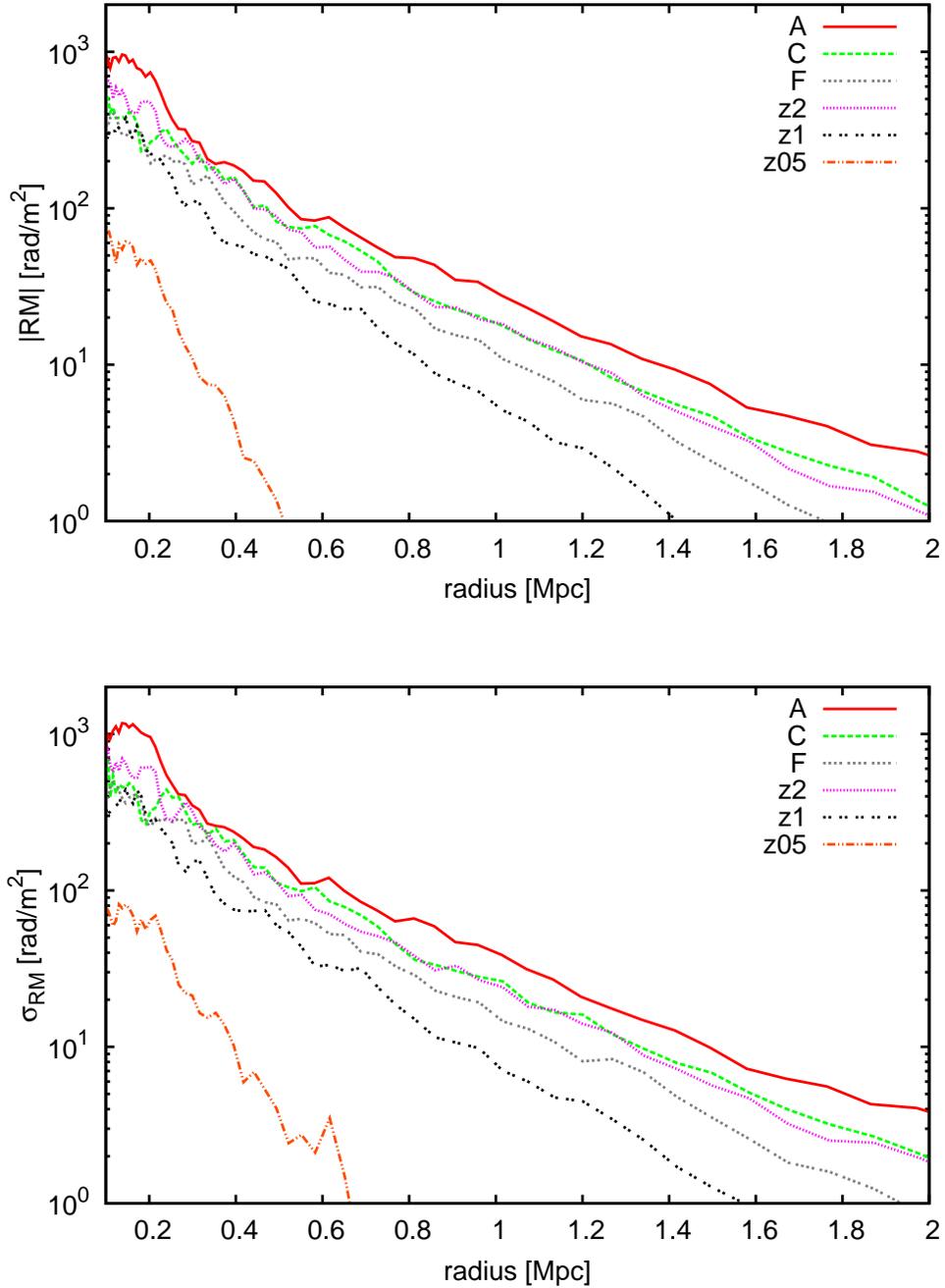,width=0.8\textwidth} 
\end{center}
 \caption{The circularly averaged radial profiles of $|RM|$ observed from y direction at $z=0$ (top)
 and the radial profiles of the standard deviation ($\sigma_{RM}$) of the same RM distribution (bottom). 
   \label{fig:rm_profile}}
\end{figure}

In Fig. \ref{fig:rm_profile}, we present the circularly averaged radial profiles of 
the absolute value of RM ($|RM|$), which is $\frac{1}{2\pi}\int^{2\pi}_{0}|RM|d\Phi$, where $\Phi$ is the azimuth angle, 
and the radial profiles of the standard deviations of RM,
which is $(\frac{1}{2\pi}\int^{2\pi}_{0}(RM-\overline{RM})^2 d\Phi)^{1/2}$, where $\overline{RM}$ = $\frac{1}{2\pi}\int^{2\pi}_{0}RM d\Phi$.
All the $|RM|$ profiles are self-similar except for z05. 
The $|RM|$ of run A are as high as 1000 rad m$^{-2}$ at the center, then drop gradually with 
increasing radius to a few rad m$^{-2}$ at 2 Mpc.  The central $|RM |$ of runs F and z1 are just above 300 rad m$^{-2}$.
The RM values are sensitive to the initial magnetic fields.
With much weaker fields, the $|RM|$ of z05 are very small and only present in the core region. 
The RM vary quickly on small scales, so their standard deviations are large. 

\begin{figure}
\begin{center}
\epsfig{file=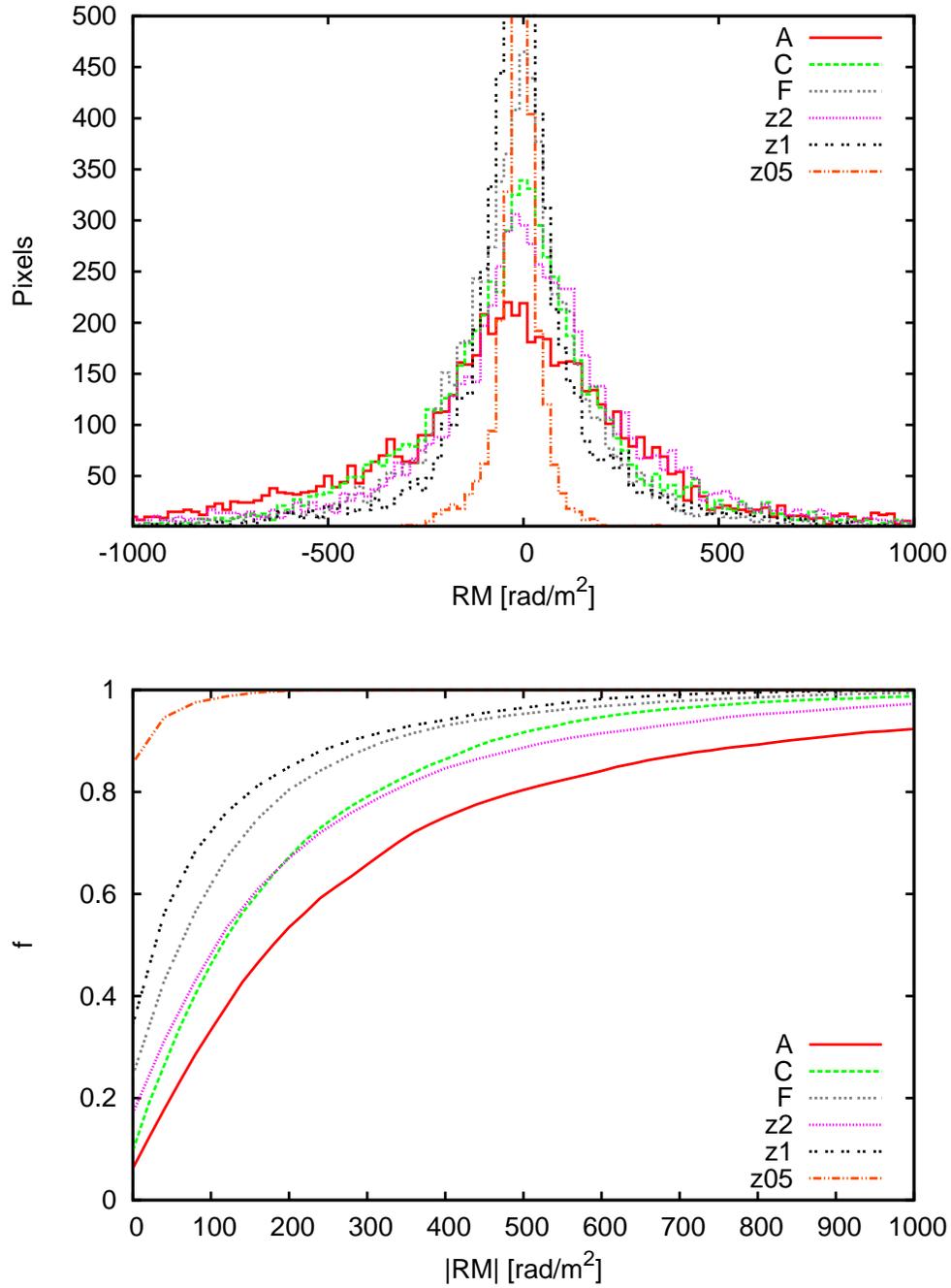,width=0.8\textwidth} 
\end{center}
 \caption{Area histogram (top) and cumulative area histogram (bottom) of RM observed from the y
 direction of the central circle of 0.5 Mpc radius at z = 0. Each pixel represents 7.81 h$^{-1}$ kpc $\times$ 7.81 h$^{-1}$ kpc.
   \label{fig:rm_histogram}}
\end{figure}

We also show the histograms of RM and the cumulative histograms of $|RM|$
inside a circle of 500 kpc in Fig \ref{fig:rm_histogram}. For the cases with magnetic fields injected at z=3, 
significant RM covers a large portion of the area. Only $\sim$ 10$\%$ of the
area is covered with RM 
close to zero for run A,  and just above 20\% for run F. The positive and negative RM are roughly symmetric. 
In run A, the average and standard deviation of RM of this area are -4.09 rad m$^{-2}$ and 307.75 rad m$^{-2}$, respectively.
There is a significant fraction of the area covered with $|RM|$ larger than 500 rad m$^{-2}$.

The general features of RM distributions of our simulations are consistent with the observations of galaxy clusters \citep[e.g.][]{Kim91, Eilek02}.
But since present observations can only measure FRM on small areas in a cluster, due to the limited number of the observed radio sources behind
or embedded in a cluster, it is too early to make a detailed comparison between the distributions of 
RM in our simulations and observations. We expect the comparison will
be available soon, when many more radio sources will be observed by EVLA.

\subsection{Importance of Simulation Resolutions on Magnetic Field Evolution}
\label{sec:comparision}

\begin{figure}
\begin{center}
\epsfig{file=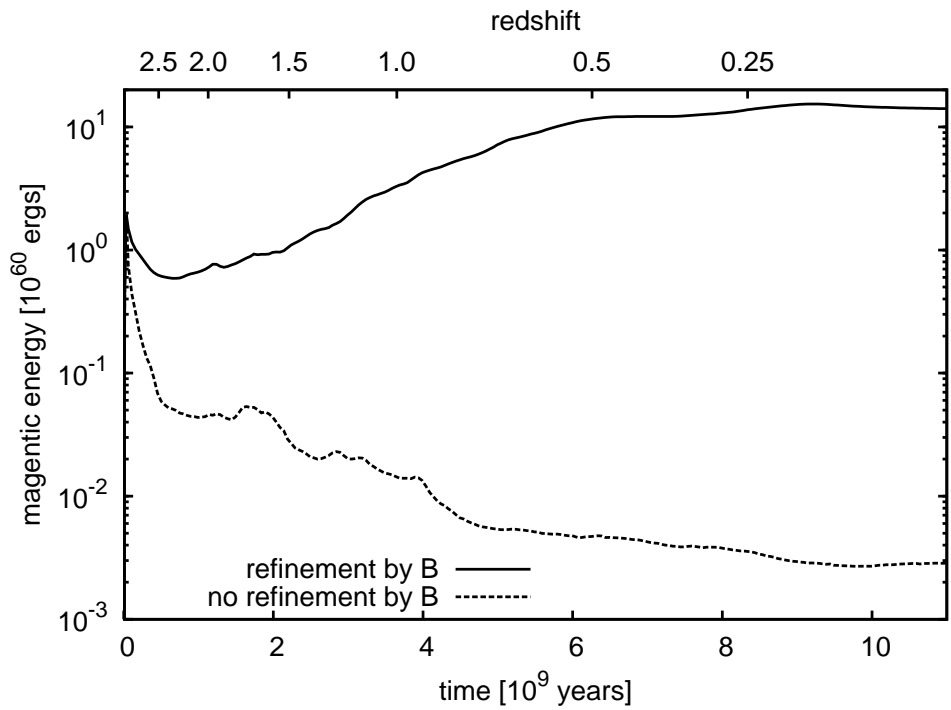,width=0.8\textwidth} 
\end{center}
 \caption{Magnetic energy evolution of the injection run A 
  with and without refinement by the magnetic field strength.
   \label{fig:MED_densityonly}}
\end{figure}

\begin{figure}
\begin{center}
\epsfig{file=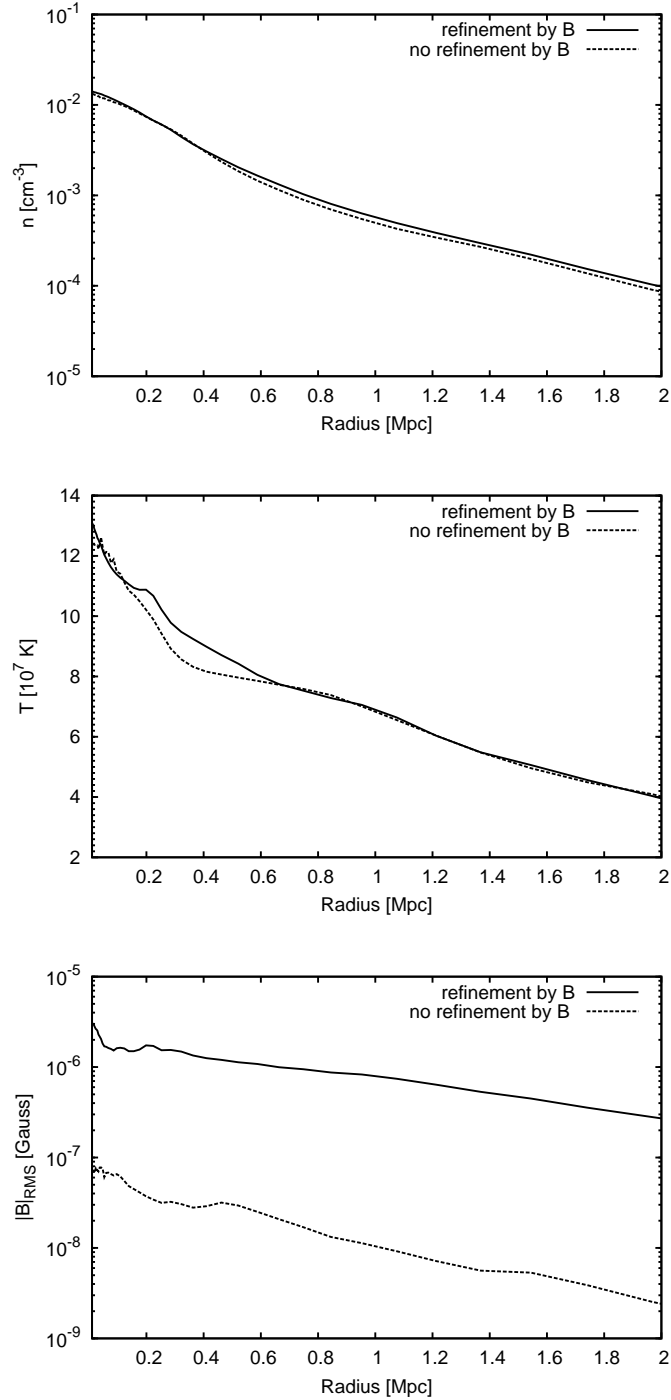,height=0.9\textheight} 
\end{center}
 \caption{Spherically averaged radial profiles of baryon density (top), temperature (middle) and RMS magnetic field
strength (bottom) at z = 0 for cases with and without refinement by the magnetic field strength.
   \label{fig:comparison_densityonly}}
\end{figure}

We find that the additional refinement criterion of magnetic field strength is very important
in following the magnetic fields correctly in our simulations. With only refinement by 
the baryon and dark matter overdensity, the spatial resolution in the outer part of the
cluster drops quickly with increasing radius. This leads to numerical 
reconnection in low resolution regions, which will destroy magnetic fields. We reran run A without the magnetic field refinement to
test the importance of resolution. 
Results are given in Fig. \ref{fig:MED_densityonly} 
and \ref{fig:comparison_densityonly}.

Without high resolution in a large volume, the simulation cannot resolve the
turbulent 
motions, and in turn, cannot capture the small-scale dynamo effect. In addition, the magnetic fields suffer much
larger numerical dissipation with the lower resolutions. The magnetic energy in
the low resolution run keeps decreasing
after injection until the end of simulation.  The few small increases in magnetic
energy that do occur correspond to major mergers.
During the early stage, the magnetic energy drops much faster in the lower
resolution run,  because most of the magnetic fields numerically reconnect in low resolution 
regions. The amplification of the magnetic field fails, most likely because the small scale turbulent motions needed by the dynamo process 
are not captured by this simulation in a large enough volume. When the simulations end at z = 0, there are 5000 times 
more magnetic energy in the case with magnetic refinement.

Though the magnetic energy evolution is very different for different refinement criteria, the cluster still forms in a similar way. 
At z = 0, the radial profiles of the baryon density and temperature are
almost the same. For density, there is only small difference out to 400 kpc due to the decrease of
resolutions in the no magnetic refinement run. The temperature profiles are different between 
200 and 600 kpc, which may be related to the differences caused by resolving the merger shocks with different resolutions. But the magnetic 
field strength profiles are totally different for these two simulations. Even in the central region with the highest resolution, the magnetic
field strength of the lower resolution case is below 0.1 $\mu$G, which is much weaker than that of the well refined one. 
Since both simulations have the same resolution in the core region, up to 50 kpc radius, the dramatic 
difference in the field strength shows that the magnetic fields amplified in the
outer regions of the cluster and transported inwards play a large role in the
magnetic field strength in the cluster core.  The radial profile of the run without
magnetic field refinement is also steeper than that of the magnetic refinement one.

This clearly shows that the simulation resolution is very important to the magnetic field evolution in the case when 
magnetic fields are initially injected in a local region. Simulations without
high resolution in a large enough region will fail to capture the small-scale
dynamo process, and lose most of the magnetic field. Whether differnet seeding processes, for example with 
initial magnetic fields everywhere, will have similar effect is still unclear. We are performing some simulations to 
address this issue. 

Though the current spatial resolution of our simulations is enough to capture the operation of the small scale dynamo in the ICM, we
don't know whether the simulation is converged. With current computational capability, we can't have simulations with even higher spatial 
resolutions in a large enough volume to test convergence. Our current results should be taken only as the lower bound of magnetic fields in the ICM. 
We will study the convergence of these simulations once the computation becomes possible.

\section{Summary}

In this work, we have described the magnetic field evolution and properties of the ICM 
with simulations in which magnetic energy was injected by an AGN outburst. We
are encouraged by the similarities seen in 
field evolution as compared to our earlier work \citep{Xu09}, which used a
completely different galaxy cluster, and magnetic energy injected at different
strengths and at different times.  Using magnetic fields from active galaxies and 
furthing amplifying them by the ICM motions is a robust way to populate the cluster with magnetic fields. 
On the other hand, since we have not verified whether the numerical convergence is reached 
in our simulations due to the limit of computational power,
the magnetic fields obtained in this paper should be treated as the minimal possible magnetic fields in the ICM.

We have simulated the AGN injection with a wide range of magnetic energy between 10$^{57}$ and 10$^{60}$ 
ergs, which covers the typical range of energy output of radio galaxies \citep{Kronberg01}.  It is interesting
that, with the huge range in injected magnetic field strength, the galaxy cluster is magnetized to a similar level, especially 
at the cluster central regions. The magnetic energy is amplified by more than 900 times in the case with the least initial  
magnetic field injection. Magnetic energy growth of some runs seems to saturate at low redshift.  
The saturated magnetic fields in the cluster center are at the micro Gauss level, 
which is consistent with observed magnetic field strengths,
as measured by different observational methods. Further study is needed to understand the saturation mechanism and why the magnetic fields saturate 
with magnetic energy at a few percent of the kinetic energy \citep{Xu09, Xu09b}. 

We also report simulations with magnetic fields injected at different redshifts. We find that magnetic fields 
injected as late as z=1 can still effectively magnetize the cluster, while the magnetic fields from AGNs after z=0.5 will
not contribute substantially to the cluster wide magnetic fields. So as long as
the magnetic fields are injected into the cluster 
before the major mergers in its formation history, the cluster will be significantly magnetized.

Since more injected magnetic energy doesn't necessarily correspond to
significantly higher final magnetic energy, additional initial magnetic fields from
multiple AGNs or from other sources (e.g. normal galaxies or early universe) may not necessarily change the 
magnetic field strength in a cluster. We have performed a simulation with two AGN injections (each injected in different halos with magnetic energy similar to run D).
Its final magnetic energy (7.62 $\times$ 10$^{60}$ ergs) and the magnetic field
profile are very similar to the single injection result. These results will be presented in a future publication.

The injected local magnetic fields mix with the ICM efficiently by the mergers and the turbulent ICM motions. 
In all cases except z05, the volume filling factors are quite high. The magnetic fields fill the central 
region of the cluster up to 1 Mpc. The magnetic field strength is not strongly correlated with the gas density, as the fields are distributed through 
a large range of gas density.

We have produced FRM maps of the whole cluster based on our simulation results. 
They have similar features as the observed cluster RM \citep[e.g.][]{Taylor93, Eilek02, Guidetti08}. 
More detailed comparisons need to wait for the additional 
observational results from the EVLA.

.

\acknowledgments
  We thank S. Colgate and H. Aluie for discussions.
  This work was supported by the LDRD and IGPP programs at LANL and by DOE/Office
  of Fusion energy Science. Computations
  were performed using the institutional computing resources at LANL.
  ENZO$\_$MHD is developed at the Laboratory for Computational Astrophysics,
  UCSD with partial support from NSF grants AST-0708960 and AST-0808184 to M.L.N.

\clearpage

\clearpage

\clearpage

\begin{thebibliography}{46}
\expandafter\ifx\csname natexlab\endcsname\relax\def\natexlab#1{#1}\fi

\bibitem[{{Bernet} {et~al.}(2008){Bernet}, {Miniati}, {Lilly}, {Kronberg}, \&
  {Dessauges-Zavadsky}}]{Bernet08}
{Bernet}, M.~L., {Miniati}, F., {Lilly}, S.~J., {Kronberg}, P.~P., \&
  {Dessauges-Zavadsky}, M. 2008, \nat, 454, 302

\bibitem[{{Boldyrev} \& {Cattaneo}(2004)}]{Boldyrev04}
{Boldyrev}, S., \& {Cattaneo}, F. 2004, Physical Review Letters, 92, 144501

\bibitem[{{Bonafede} {et~al.}(2010){Bonafede}, {Feretti}, {Murgia}, {Govoni},
  {Giovannini}, {Dallacasa}, {Dolag}, \& {Taylor}}]{Bonafede10}
{Bonafede}, A., {Feretti}, L., {Murgia}, M., {Govoni}, F., {Giovannini}, G.,
  {Dallacasa}, D., {Dolag}, K., \& {Taylor}, G.~B. 2010, \aap, 513, A30

\bibitem[{{Brandenburg} \& {Subramanian}(2005)}]{Brandenburg05}
{Brandenburg}, A., \& {Subramanian}, K. 2005, \physrep, 417, 1

\bibitem[{{Burbidge}(1959)}]{Burbidge59}
{Burbidge}, G.~R. 1959, \apj, 129, 849

\bibitem[{{Carilli} \& {Taylor}(2002)}]{Carilli02}
{Carilli}, C.~L., \& {Taylor}, G.~B. 2002, \araa, 40, 319

\bibitem[{{Cho} \& {Vishniac}(2000)}]{Cho00}
{Cho}, J., \& {Vishniac}, E. T. 2000, \apj, 538, 217

\bibitem[{{Colgate} \& {Li}(2000)}]{Colgate00}
{Colgate}, S.~A., \& {Li}, H. 2000, in IAU Symposium, Vol. 195, Highly
  Energetic Physical Processes and Mechanisms for Emission from Astrophysical
  Plasmas, ed. P.~C.~H. {Martens}, S.~{Tsuruta}, \& M.~A. {Weber}, 255--264

\bibitem[{{Collins} {et~al.}(2010){Collins}, {Xu}, {Norman}, {Li}, \&
  {Li}}]{Collins09}
{Collins}, D.~C., {Xu}, H., {Norman}, M.~L., {Li}, H., \& {Li}, S. 2010, \apjs, 186, 308

\bibitem[{{Croston} {et~al.}(2005){Croston}, {Hardcastle}, {Harris}, {Belsole},
  {Birkinshaw}, \& {Worrall}}]{Croston05}
{Croston}, J.~H., {Hardcastle}, M.~J., {Harris}, D.~E., {Belsole}, E.,
  {Birkinshaw}, M., \& {Worrall}, D.~M. 2005, \apj, 626, 733

\bibitem[{{Dolag} {et~al.}(2002){Dolag}, {Bartelmann}, \& {Lesch}}]{Dolag02}
{Dolag}, K., {Bartelmann}, M., \& {Lesch}, H. 2002, \aap, 387, 383

\bibitem[{{Donnert} {et~al.}(2009){Donnert}, {Dolag}, {Lesch}, \&
  {M{\"u}ller}}]{Donnert08}
{Donnert}, J., {Dolag}, K., {Lesch}, H., \& {M{\"u}ller}, E. 2009, \mnras, 392,
  1008

\bibitem[{{Dubois} {et~al.}(2009){Dubois}, {Devriendt}, {Slyz}, \&
  {Silk}}]{Dubois09}
{Dubois}, Y., {Devriendt}, J., {Slyz}, A., \& {Silk}, J. 2009, \mnras, 399, L49

\bibitem[{{Dubois} \& {Teyssier}(2008)}]{Dubois08}
{Dubois}, Y., \& {Teyssier}, R. 2008, \aap, 482, L13

\bibitem[{{Eilek} \& {Owen}(2002)}]{Eilek02}
{Eilek}, J.~A., \& {Owen}, F.~N. 2002, \apj, 567, 202

\bibitem[{{Eisenstein} \& {Hu}(1999)}]{Eisenstein99}
{Eisenstein}, D.~J., \& {Hu}, W. 1999, \apj, 511, 5

\bibitem[{{En{\ss}lin} \& {Vogt}(2006)}]{Ensslin06}
{En{\ss}lin}, T.~A., \& {Vogt}, C. 2006, \aap, 453, 447

\bibitem[{{Fan} {et~al.}(2001){Fan}, {Strauss}, {Schneider}, {Gunn}, {Lupton},
  {Becker}, {Davis}, {Newman}, {Richards}, {White}, {Anderson}, {Annis},
  {Bahcall}, {Brunner}, {Csabai}, {Hennessy}, {Hindsley}, {Fukugita}, {Kunszt},
  {Ivezi{\'c}}, {Knapp}, {McKay}, {Munn}, {Pier}, {Szalay}, \& {York}}]{Fan01}
{Fan}, X., {Strauss}, M.~A., {Schneider}, D.~P., {Gunn}, J.~E., {Lupton},
  R.~H., {Becker}, R.~H., {Davis}, M., {Newman}, J.~A., {Richards}, G.~T.,
  {White}, R.~L., {Anderson}, Jr., J.~E., {Annis}, J., {Bahcall}, N.~A.,
  {Brunner}, R.~J., {Csabai}, I., {Hennessy}, G.~S., {Hindsley}, R.~B.,
  {Fukugita}, M., {Kunszt}, P.~Z., {Ivezi{\'c}}, {\v Z}., {Knapp}, G.~R.,
  {McKay}, T.~A., {Munn}, J.~A., {Pier}, J.~R., {Szalay}, A.~S., \& {York},
  D.~G. 2001, \aj, 121, 54

\bibitem[{{Feretti}(1999)}]{Feretti99}
{Feretti}, L. 1999, in Diffuse Thermal and Relativistic Plasma in Galaxy
  Clusters, ed. H.~{Boehringer}, L.~{Feretti}, \& P.~{Schuecker}, 3--8

\bibitem[{{Ferrari} {et~al.}(2008){Ferrari}, {Govoni}, {Schindler}, {Bykov}, \&
  {Rephaeli}}]{Ferrari08}
{Ferrari}, C., {Govoni}, F., {Schindler}, S., {Bykov}, A.~M., \& {Rephaeli}, Y.
  2008, Space Sci. Rev., 134, 93

\bibitem[{{Furlanetto} \& {Loeb}(2001)}]{Furlanetto01}
{Furlanetto}, S.~R., \& {Loeb}, A. 2001, \apj, 556, 619

\bibitem[{{Giovannini} {et~al.}(2009){Giovannini}, {Bonafede}, {Feretti},
  {Govoni}, {Murgia}, {Ferrari}, \& {Monti}}]{Giovanini09}
{Giovannini}, G., {Bonafede}, A., {Feretti}, L., {Govoni}, F., {Murgia}, M.,
  {Ferrari}, F., \& {Monti}, G. 2009, \aap, 507, 1257

\bibitem[{{Govoni} {et~al.}(2006){Govoni}, {Murgia}, {Feretti}, {Giovannini},
  {Dolag}, \& {Taylor}}]{Govoni06}
{Govoni}, F., {Murgia}, M., {Feretti}, L., {Giovannini}, G., {Dolag}, K., \&
  {Taylor}, G.~B. 2006, \aap, 460, 425

\bibitem[{{Guidetti} {et~al.}(2008){Guidetti}, {Murgia}, {Govoni}, {Parma},
  {Gregorini}, {de Ruiter}, {Cameron}, \& {Fanti}}]{Guidetti08}
{Guidetti}, D., {Murgia}, M., {Govoni}, F., {Parma}, P., {Gregorini}, L., {de
  Ruiter}, H.~R., {Cameron}, R.~A., \& {Fanti}, R. 2008, \aap, 483, 699

\bibitem[{{Haugen} et al. (2004)}]{Haugen04}
{Haugen},  N. E. L., {Brandenbury}, A., \& {Dobler}, W. 2004, Phys. Rev. E, 70, 016308

\bibitem[{{Kim} {et~al.}(1991){Kim}, {Kronberg}, \& {Tribble}}]{Kim91}
{Kim}, K.-T., {Kronberg}, P.~P., \& {Tribble}, P.~C. 1991, \apj, 379, 80

\bibitem[{{Kronberg} {et~al.}(2008){Kronberg}, {Bernet}, {Miniati}, {Lilly},
  {Short}, \& {Higdon}}]{Kronberg08}
{Kronberg}, P.~P., {Bernet}, M.~L., {Miniati}, F., {Lilly}, S.~J., {Short},
  M.~B., \& {Higdon}, D.~M. 2008, \apj, 676, 70

\bibitem[{{Kronberg} {et~al.}(2001){Kronberg}, {Dufton}, {Li}, \&
  {Colgate}}]{Kronberg01}
{Kronberg}, P.~P., {Dufton}, Q.~W., {Li}, H., \& {Colgate}, S.~A. 2001, \apj,
  560, 178

\bibitem[{{Kulsrud} {et~al.}(1997){Kulsrud}, {Cen}, {Ostriker}, \&
  {Ryu}}]{Kulsrud97}
{Kulsrud}, R.~M., {Cen}, R., {Ostriker}, J.~P., \& {Ryu}, D. 1997, \apj, 480,
  481

\bibitem[{{Li} {et~al.}(2006){Li}, {Lapenta}, {Finn}, {Li}, \&
  {Colgate}}]{Li06}
{Li}, H., {Lapenta}, G., {Finn}, J.~M., {Li}, S., \& {Colgate}, S.~A. 2006,
  \apj, 643, 92

\bibitem[{{McNamara} \& {Nulsen}(2007)}]{McNamara07}
{McNamara}, B.~R., \& {Nulsen}, P.~E.~J. 2007, \araa, 45, 117

\bibitem[{{McNamara} {et~al.}(2005){McNamara}, {Nulsen}, {Wise}, {Rafferty},
  {Carilli}, {Sarazin}, \& {Blanton}}]{McNamara05}
{McNamara}, B.~R., {Nulsen}, P.~E.~J., {Wise}, M.~W., {Rafferty}, D.~A.,
  {Carilli}, C., {Sarazin}, C.~L., \& {Blanton}, E.~L. 2005, \nat, 433, 45

\bibitem[{{Motl} {et~al.}(2004){Motl}, {Burns}, {Loken}, {Norman}, \&
  {Bryan}}]{Motl04}
{Motl}, P.~M., {Burns}, J.~O., {Loken}, C., {Norman}, M.~L., \& {Bryan}, G.
  2004, \apj, 606, 635

\bibitem[{{Nagai} {et~al.}(2007){Nagai}, {Vikhlinin}, \& {Kravtsov}}]{Nagai07}
{Nagai}, D., {Vikhlinin}, A., \& {Kravtsov}, A.~V. 2007, \apj, 655, 98

\bibitem[{{Nakamura} {et~al.}(2006){Nakamura}, {Li}, \& {Li}}]{Nakamura06}
{Nakamura}, M., {Li}, H., \& {Li}, S. 2006, \apj, 652, 1059

\bibitem[{{Nulsen} {et~al.}(2005){Nulsen}, {McNamara}, {Wise}, \&
  {David}}]{Nulsen05}
{Nulsen}, P.~E.~J., {McNamara}, B.~R., {Wise}, M.~W., \& {David}, L.~P. 2005,
  \apj, 628, 629

\bibitem[{{Roettiger} {et~al.}(1999){Roettiger}, {Stone}, \&
  {Burns}}]{Roettiger99}
{Roettiger}, K., {Stone}, J.~M., \& {Burns}, J.~O. 1999, \apj, 518, 594

\bibitem[{{Ryu} {et~al.}(2008){Ryu}, {Kang}, {Cho}, \& {Das}}]{Ryu08}
{Ryu}, D., {Kang}, H., {Cho}, J., \& {Das}, S. 2008, Science, 320, 909

\bibitem[{{Schekochihin} \& {Cowley}(2007)}]{Schekochihin07}
{Schekochihin}, A., \& {Cowley}, S. 2007, in Magnetohydrodynamics - Historical Evolution and Trends, ed. S. Molokov, R. Moreau, \& H. Moffatt 
(Berlin: Springer) 85

\bibitem[{{Spergel} {et~al.}(2007){Spergel}, {Bean}, {Dor{\'e}}, {Nolta},
  {Bennett}, {Dunkley}, {Hinshaw}, {Jarosik}, {Komatsu}, {Page}, {Peiris},
  {Verde}, {Halpern}, {Hill}, {Kogut}, {Limon}, {Meyer}, {Odegard}, {Tucker},
  {Weiland}, {Wollack}, \& {Wright}}]{Spergel07}
{Spergel}, D.~N., {Bean}, R., {Dor{\'e}}, O., {Nolta}, M.~R., {Bennett}, C.~L.,
  {Dunkley}, J., {Hinshaw}, G., {Jarosik}, N., {Komatsu}, E., {Page}, L.,
  {Peiris}, H.~V., {Verde}, L., {Halpern}, M., {Hill}, R.~S., {Kogut}, A.,
  {Limon}, M., {Meyer}, S.~S., {Odegard}, N., {Tucker}, G.~S., {Weiland},
  J.~L., {Wollack}, E., \& {Wright}, E.~L. 2007, \apjs, 170, 377

\bibitem[{{Stone} et al. (1998)}]{Stone98}
{Stone}, J. M., {Ostriker}, E. C., \& {Gammie}, C. F. 1998, \apj, 508, L99

\bibitem[{{Subramanian} {et~al.}(2006){Subramanian}, {Shukurov}, \&
  {Haugen}}]{Subramanian06}
{Subramanian}, K., {Shukurov}, A., \& {Haugen}, N.~E.~L. 2006, \mnras, 366,
  1437

\bibitem[{{Sunyaev} et al. (2003)}]{Sunyaev03}
{Sunyaev}, R.~A., {Norman}, M.~L., {Bryan}, G.~L. 2003, Astronomy Letters, 29, 783

\bibitem[{{Taylor} \& {Perley}(1993)}]{Taylor93}
{Taylor}, G.~B., \& {Perley}, R.~A. 1993, \apj, 416, 554

\bibitem[{Turk(2008)}]{Turk08}
Turk, M. 2008, in Proceedings of the 7th Python in Science Conference, ed.
  G.~Varoquaux, T.~Vaught, \& J.~Millman, Pasadena, CA USA, 46 -- 50

\bibitem[{{Vazza} {et~al.}(2009){Vazza}, {Brunetti}, {Kritsuk}, {Wagner},
  {Gheller}, \& {Norman}}]{Vazza09}
{Vazza}, F., {Brunetti}, G., {Kritsuk}, A., {Wagner}, R., {Gheller}, C., \&
  {Norman}, M. 2009, \aap, 504, 33

\bibitem[{{Vogt} \& {En{\ss}lin}(2003)}]{Vogt03}
{Vogt}, C., \& {En{\ss}lin}, T.~A. 2003, \aap, 412, 373

\bibitem[{{Voit}(2005)}]{Voit05}
{Voit}, G.~M. 2005, Reviews of Modern Physics, 77, 207

\bibitem[{{Xu}(2009)}]{Xu09b}
{Xu}, H. 2009, PhD thesis, University of California, San Diego

\bibitem[{{Xu} {et~al.}(2008){Xu}, {Li}, {Collins}, {Li}, \& {Norman}}]{Xu08a}
{Xu}, H., {Li}, H., {Collins}, D., {Li}, S., \& {Norman}, M.~L. 2008, \apjl,
  681, L61

\bibitem[{{Xu} {et~al.}(2009){Xu}, {Li}, {Collins}, {Li}, \& {Norman}}]{Xu09}
{Xu}, H., {Li}, H., {Collins}, D.~C., {Li}, S., \& {Norman}, M.~L. 2009, \apjl,
  698, L14

\end{thebibliography}
\end{document}